\documentclass[12pt,a4,sans]{article}
\usepackage[latin1]{inputenc}
\usepackage{amssymb}
\usepackage{amsmath}
\usepackage{amsthm}
\usepackage{amsfonts}
\usepackage[pdftex]{graphicx}
\usepackage{geometry}
\usepackage{braket}
\usepackage{enumerate}
\usepackage[usenames,dvipsnames]{xcolor}
\usepackage{setspace}
\usepackage[backgroundcolor=white,linecolor=black]{todonotes}
\usepackage[in]{fullpage}
\usepackage{array}
\usepackage{wrapfig}
\usepackage[font=small,labelfont=bf]{caption}
\usepackage{anysize}
\usepackage{float}
\usepackage{graphicx}
\usepackage{xcolor}
\usepackage{mathtools}
\usepackage{youngtab}
\newcommand{\be}{\begin{equation}}
\newcommand{\ee}{\end{equation}}
\newcommand{\bea}{\begin{eqnarray}\displaystyle}
\newcommand{\eea}{\end{eqnarray}}

\makeatletter
\@addtoreset{equation}{section}
\makeatother
\renewcommand{\theequation}{\thesection.\arabic{equation}}
\def\one{{\hbox{ 1\kern-.8mm l}}}
\def\zero{{\hbox{ 0\kern-1.5mm 0}}}

\def\mC{ \mathbb{C}}

\def\cA{{\cal A}} \def\cB{{\cal B}} 
  \def\cF{{\cal F}}
  
 \def\cK{{\cal K}} 
\def\cM{{\cal M}} \def\cN{{\cal N}} \def\cO{{\cal O}}
  
 \def\cT{{\cal T}} 
  
 \def\cZ{{\cal Z}}
\newcommand{\comments}[1]{}
\newcommand{\la}{\langle}
\newcommand{\ra}{\rangle}

\newcommand{\A}{\mathcal{A}}

\newcommand{\Z}{\mathcal{Z}}
\newcommand{\N}{\mathcal{N}}

\newcommand{\Tr}{\text{Tr}}

\newcommand{\nn}{\nonumber\\[3mm]}

\newcommand{\lara}[1]{\left\langle #1 \right\rangle}
\def\mA{ \mathbb{A}} 
\def\mB{ \mathbb{B}}

\begin{document}

\makeatletter
\@addtoreset{equation}{section}
\makeatother
\renewcommand{\theequation}{\thesection.\arabic{equation}}

\rightline{QMUL-PH-15-25}
\vspace{1.8truecm}

\vspace{15pt}

%%%%%%%%%%%%%%%%%

{\LARGE{ 
\centerline{\bf Permutation Centralizer Algebras } 
\centerline{ \bf and Multi-Matrix Invariants } 
}} 

\vskip.5cm 

\thispagestyle{empty} \centerline{
 {\large \bf Paolo Mattioli
 \footnote{ {\tt p.mattioli@qmul.ac.uk}} }
 {\large \bf and Sanjaye Ramgoolam
 \footnote{ {\tt s.ramgoolam@qmul.ac.uk}} }}
 
\setcounter{footnote}{0} 
 
\vspace{.4cm}
\centerline{{\it Centre for Research in String Theory, School of Physics and Astronomy},}
\centerline{{ \it Queen Mary University of London},} \centerline{{\it
 Mile End Road, London E1 4NS, UK}}

\vspace{1truecm}

%%%%%%%%%%%%%%%%%
\thispagestyle{empty}

\centerline{\bf ABSTRACT}

\vskip.2cm 
We introduce a class of permutation centralizer algebras which underly the combinatorics of multi-matrix gauge invariant observables. One family of such non-commutative algebras is parametrised by two integers. Its Wedderburn-Artin decomposition explains the counting of restricted Schur operators, which were introduced in the physics literature to describe open strings attached to giant gravitons and were subsequently used to diagonalize the Gaussian inner product for gauge invariants of 2-matrix models. The structure of the algebra, notably its dimension, its centre and its maximally commuting sub-algebra, is related to Littlewood-Richardson numbers for composing Young diagrams. It gives a precise characterization of the minimal set of charges needed to distinguish arbitrary matrix gauge invariants, which are related to enhanced symmetries in gauge theory. The algebra also gives a star product for matrix invariants. The centre of the algebra allows efficient computation of a sector of multi-matrix correlators. These generate the counting of a certain class of bi-coloured ribbon graphs with arbitrary genus. 

\vskip.2cm 

\newpage
%\pagenumbering{roman}
\tableofcontents

\pagebreak
\section{Introduction }

A number of questions on gauge invariant functions and correlators of multiple-matrices 
have been studied in the context of $\mathcal{N}=4$ Super Yang-Mills (SYM). The impetus for these developments in physics has come from the AdS/CFT correspondence
\cite{Maldacena:1997re,Witten:1998qj,Gubser:1998bc}, notably the duality between the $\N=4$ SYM theory with $U(N)$ gauge group and $ AdS_5 \times S^5 $. Local composite operators are $U(N)$ gauge invariants. CFT gives extra motivation because of the operator-state correspondence. Quantum states correspond to local operators, which are composite fields. These can be matrix-valued fields which are space-time scalars, fermions, field strengths or covariant derivatives of these. A generic problem is to understand $ U(N)$ invariants constructed from a number $n$ of such fields
\bea\label{initial} 
{ \cF }_{ 1, i_1 }^{ j_1 } \cdots {\cF }_{n , i_n}^{ j_n} 
\eea
This is subsequently used to understand their correlation functions. 
The $n$ upper indices each transform in the fundamental of $U(N)$ while the lower indices transform in the anti-fundamental. Hence, an important ingredient is 
the nature of the invariants in 
\bea 
V^{ \otimes n } \otimes \bar V^{ \otimes n } 
\eea
The number of linearly independent invariants is $n!$. They are obtained by multiplying (\ref{initial}) with a product of $n$ Kronecker delta functions, contracted 
with a permutation $ \sigma \in S_n$. As $n$ varies, we are interested in all possible values of $n$, so the properties of 
$$
\mathbb {C }[ S_{ \infty} ] = \bigoplus_{ n = 0 }^{ \infty} \mathbb{ C } [S_{ n } ]
$$ 
become important. If all the $n$ operators are the same \emph{e.g.} a complex matrix $ X = \phi_1 + i \phi_2$ where $ \phi_1 , \phi_2$ are two of the six hermitian matrices transforming in the vector of $ SO(6)$, then the invariants are multi-traces, of which there are $p(n)$, the number of partitions of $n$. In terms of the permutations, the composite operators are 
\bea\label{opclass} 
\cO_{ \sigma } (X ) = X^{ i_1}_{ i_{ \sigma (1)} } \cdots X^{ i_n }_{ i_{\sigma (n)} } 
\eea
Distinct $ \sigma $ related by conjugation, i.e. $ \sigma $ and $ \gamma \sigma \gamma^{-1} $ for some $ \gamma \in S_n$ 
give the same operator
\bea 
\cO_{ \sigma } ( X ) = \cO_{ \gamma \sigma \gamma^{-1} } ( X ) 
\eea
When we consider invariants built from two types of matrices, say $m$ copies of $X$ 
and $n$ copies of $Y$, then we encounter equivalence classes 
\bea 
\sigma \sim \gamma \sigma \gamma^{-1} 
\eea
where $ \sigma \in S_{ m+n } $ and $ \gamma \in S_m \times S_n$. 

The fact that the enumeration of gauge invariant operators can be effectively done by using a formulation in terms of equivalence classes of permutations has driven significant progress in the construction of operators and computation of correlators for the half-BPS sector, the perturbations of the half-BPS operators as well as quarter BPS operators. 
Two key facts have been used. One is that, by using the Fourier transformation 
which relates functions on a group to matrix elements of irreducible representations, 
nice orthogonal bases of functions on these equivalence classes can be found. 
In mathematics, in the context of compact groups this is known as the Peter-Weyl theorem. 
In the context of finite groups, this follows from the Schur orthogonality relations. 
This leads to the construction of operators in the half-BPS sector parametrised 
by Young diagrams \cite{CJR01,CR02}. For the two-matrix sector, one application of this thinking leads to {\it restricted Schur operators}. These are labelled by three young diagrams and a pair of multiplicity labels: a Young diagram $R_1$ with $m$ boxes, a Young 
diagram $ R_2$ with $n$ boxes and a third diagram $R$ with $m+n$ boxes. The two multiplicity labels each run over a space of dimension equal to $ g (R_1 , R_2 ; R ) $, which is equal to the Littlewood-Richardson (LR) coefficient for the number of times $R$ appears in the tensor product of $ R_1 \otimes R_2$ \cite{BHLN02,BBFH04,KSS1,KSS2,BKS}. LR coefficients will be reviewed as needed in this paper (see Appendix \ref{Appendix: LR rule for hooks}).

One reason for the efficacy of permutation groups in enumeration of gauge invariant operators is Schur-Weyl duality. This states that the tensor product of $n$ copies of the fundamental of $U(N)$ decomposes into a direct sum of irreps of $ S_n \times U(N)$ 
\bea 
V_N^{ \otimes n } = \bigoplus_{ \substack { R \vdash N \\ c_1 (R ) \le N } } V_R^{S_n} \otimes V_R^{ U(N)} 
\eea
Each summand is labelled by a Young diagram, and the Young diagrams are 
constrained to have no more than $N$ rows, equivalently the first column $c_1(R)$ 
is no greater than $N$. This uses the fact that Young diagrams are used to classify 
representations of $S_n$ as well as representations of $U(N)$. This is useful in the permutation approach to gauge invariant operators, because it says that once we have organised operators according to representation data for $S_n$, it is easy to implement finite $N$ constraints. In the one-matrix problem, the single Young diagram label $R$ is cut-off at $ N$, $ c_1 ( R)\le N $. This leads directly to the connection between the stringy exclusion principle for giant gravitons and Young diagrams \cite{MS98,MST00,BBNS02,CJR01}. In the two-matrix problem, the Young diagram $R$ is cut-off at $ c_1 ( R)\le N $, which implies cut-offs for $R_1 , R_2$. 
The 2-matrix problem can also be approached using the walled Brauer algebra  $ B_N ( m, n ) $ and its representation theory \cite{KR1}. A third way to enumerate two-matrix invariants, also based on permutations but involving Clebsch-Gordan multiplicities of $S_n$, keeps  the $U(2)$ global symmetry  manifest \cite{BHR1,BHR2}. 

Aside from enumerating gauge invariant operators, the permutation structures have been used to compute correlators. Correlators in free field theory are obtained by sums over Wick contractions. These sums are themselves parametrised by permutations. Correlators of gauge invariant operators are thus given in terms of these Wick permutations and the permutations which enumerate the operators. Hence there are elegant formulae for the correlation functions in terms of permutations. It can be shown that the 2-point functions of gauge invariant operators in the 2-matrix sector are diagonalized by operators constructed using representation bases. 
This was done with the Brauer basis in \cite{KR1}, with the $U(2)$ covariant basis in \cite{BHR1,BHR2} and with the restricted Schur 
basis in \cite{BCK08,scollins08}. The restricted Schur and covariant basis results have been extended beyond $\N=4$ SYM to the sector of holomorphic operators in general quiver gauge theories \cite{CDK13,KMMP12,PR11,quivcalc,QW15,KKN14}
which have been shown to include sectors related to generalized oscillators \cite{Berenstein:2015ooa}. Aspects involving Frobenius algebras 
have been studied in \cite{Kimura:2014mka}. Within $\N=4 $ SYM, perturbations of half-BPS giant graviton operators have been studied and integrability at one-loop \cite{KR12,CKH11,KMS11,KDGM11,KR12,KKN14} and beyond has been established.

As a way to understand the existence of the different bases in the multi-matrix problems, the paper \cite{KR2} conducted a detailed study of enhanced symmetries in the free limit of Yang Mills theories. The authors showed that Casimir-like elements constructed from Noether charges of these enhanced symmetries can be used to understand these different bases. Different sets of these Casimir-like charges each consist of mutually commuting simultaneously diagonalizable operators, which associate the labels of the basis with eigenvalues of Casimir-like charges. Thus there is a set of Casimir-like elements for the restricted Schur basis, another set for the covariant basis and yet another set for the Brauer basis. The enhanced symmetries themselves take the form of products of unitary groups, but the action of these Casimirs on gauge invariant operators can be related, through applications of Schur-Weyl duality, to the algebraic structure of certain algebras constructed from the equivalence classes of permutations or of Brauer algebra elements discussed above. The discussion of charges which identify matrix invariants for general classical groups has been given using a different approach in \cite{Diaz:2014ixa}. While a uniform treatment of the Young diagram labels has been achieved, a treatment of the multiplicity labels running over Littlewood-Richardson coefficients in that approach remains 
 an interesting open problem. 

This paper was motivated by the goal of obtaining a systematic understanding of the algebraic structures involved in the  construction of charges in \cite{KR2}. 
To be more precise, we will define the notion of {\it permutation centralizer algebras}. A particular class of these, denoted as $ \cA ( m ,n ) $, will be our main focus. 
Many of the important formulae we will use have already appeared in the physics literature. 
Nevertheless the $ \cA ( m ,n ) $, as associative algebras with non-degenerate pairing, have not been made fully explicit. This paper proposes that these algebras are interesting to study intrinsically, disentangled from the contingencies of being embedded in a bigger symmetric group algebra, their simplicity hidden among the application to matrix correlators for matrices of size $N$. 
Here we define the algebras $ \cA ( m ,n ) $, study their structure, and subsequently describe how they are relevant to matrix theory invariants. We expect that a deeper study of this algebraic structure has the potential to give a lot of information about correlators in free Yang-Mills theory, in the loop corrected theory, at all orders in the $1/N$ expansion. This paper is a step in this direction. Much as it is valuable to abstract Riemannian geometry from the study of submanifolds of Euclidean spaces, abstracting a family of algebras intrinsic to permutations hidden in the mathematics of matrix theory should be fruitful.

We describe the organization of the paper. In section \ref{sec:def} we introduce the definition of permutation centralizer algebras. We consider four key examples of these algebras, which are useful in the context of gauge-invariant operators. 
In section \ref{sec:struct}, we focus on the algebras $ \cA ( m ,n ) $ formed by equivalence classes of permutations in $ S_{ m +n } $, with equivalence generated by conjugation with permutations in $ S_m \times S_n$. The dimension of this algebra is
\begin{align}
|\A(m,n)|=\sum_{ R_1 \vdash m , R_2 \vdash n\atop R\vdash m+n } g ( R_1 , R_2 ; R )^2 
\end{align}
where $g ( R_1 , R_2 ; R )$ is the LR coefficient for the triplet of Young diagram $(R_1,R_2,R)$ made with $(m,n,m+n)$ boxes respectively.
We will show that this is an associative algebra with a non-degenerate pairing. As a result, we know from the Wedderburn-Artin theorem that it is isomorphic to a direct sum of matrix algebras $\mathcal { M} at $ \cite{goodman2000representations,RamChap1}:
\begin{align}
\A(m,n)=\bigoplus_{a}\cM{at}_a
\end{align}
In eq. \eqref{Wedderburn theorem} we give a more precise version of this formula, where the index $a$ is identified with triplets $(R_1,R_2,R)$ with non-vanishing LR coefficient $g(R_1,R_2;R)$.
The construction of restricted Schur operators in gauge theory is used to give the Wedderburn-Artin decomposition of $ \cA ( m ,n )$. Two sub-algebras will be of interest. The centre of the algebra $ \cZ ( m ,n ) $ is the subspace of the algebra which commutes with any element of $ \cA( m ,n ) $. The dimension of this centre is equal to the number of triples $ ( R_1 , R_2 , R )$ of Young diagrams, with numbers of boxes equal to $ ( m , n , m+n )$, for which the LR coefficient is non-zero. It is useful to develop some formulae for the non-degenerate pairing on the centre, using characters of $ S_{m+n}, S_m , S_n$. The Wedderburn-Artin decomposition also highlights the importance of a maximally commuting sub-algebra $ \cM ( m , n ) $. The dimension of this sub-algebra is the sum of Littlewood-Richardson coefficients $ g ( R_1, R_2 ; R)$. Appendix \ref{sec:AnalDim} gives a multi-variable generating function for this sum of LR coefficients. We explain the relevance of the this sub-algebra to the enhanced symmetry charges studied in \cite{EHS}. In particular we give a precise algebraic characterization \eqref{charge-min} for the \emph{minimal number of charges needed to identify all 2-matrix gauge-invariant operators}. The evaluation of this number is an open problem for the future.

In section \ref{Section: Star product}, we explain some further physical implications of the 
permutation centralizer algebras. The simplest of these algebras is the algebra of 
class sums of permutations. Given the one-to-one correspondence between matrix operators 
and conjugacy classes of permutations given in (\ref{opclass}), this means that 
there is a corresponding product on half-BPS operators. This is not the usual 
product obtained by multiplying the gauge invariant operator built from $X$ under which 
the dimension of the operator adds. The product on the class sums rather gives a
product for the BPS operators of fixed dimension, a product which is associative and 
admits a non-degenerate pairing. We will refer to this as a \emph{star product for half-BPS operators}. 
We explain the relevance of this star product for the computation of correlators. 
Similarly the product on the algebra $ \cA ( m , n ) $ gives a \emph{ star product for 
gauge invariant polynomials in two matrices}, with degree $m$ in the $X$'s and 
degree $n$ in the $Y$'s. In the physics application, there is a closed associative star product 
on the space of quarter-BPS operators at zero Yang-Mills coupling. 
Conversely the usual product of gauge invariants gives a product on $ \cA ( \infty ,\infty ) $
\bea 
 \cA ( \infty ,\infty ) = \bigoplus_{ m , n = 0 }^{ \infty} \cA ( m , n ) 
\eea
which is the direct sum over all $ m , n$. Thus $ \cA ( \infty ,\infty ) $ has two products 
one of which closes at fixed $m , n $. This generalizes a structure seen in the study of symmetric polynomials. 

%( MAYBE - say that for $ \cS ( \infty ) $ the two products are known 
 % from theory of symmetric polynomials. Also mention references to russian literature related to these 2 products- probably in conclusion) 

In section \ref{sec:finiteNcorrelator}, we show that the study of the structure of the 
algebra $ \cA ( m ,n ) $ we developed in section \ref{sec:struct} is useful for the computation of correlators of 2-matrix gauge invariants. In particular, we identify an \emph{ efficiently computable sector of {\it central } gauge invariant operators} whose correlators can be computed using the knowledge of characters of $S_{ m+n} , S_m , S_n $. It does not require the knowledge of more detailed data such as matrix elements $ D^R_{ i j } ( \sigma ) $ or branching coefficients for $ S_{ m +n } \rightarrow S_{ m } \times S_n $. To illustrate the simplicity of this central sector, we compute the two-point function 
\bea 
\lara{ \Tr ( X^m Y^n ) \Tr ( ( X^{ \dagger} )^m ( Y^{ \dagger} )^n ) }
\eea
at finite $N$. The computation requires a calculation of Littlewood-Richardson coefficients 
$ g ( R_1 , R_2 ; R)$ where $R_1, R_2$ are hook-shaped Young diagrams. This computation is given in Appendix \ref{Appendix: LR rule for hooks}. Further technical aspects of the computation are given in Appendix \ref{Appendix: two point generating function}. The computation agrees with the one in \cite{Bhattacharyya:2008xy} 
which was done with explicit Young-Yamanouchi symbols which can be used to construct states in irreps $R$ and describe their reduction to $R_1, R_2$. 

In section \ref{conclusions}, we outline some future research directions related to the present results.

\section{Definitions and Key examples }\label{sec:def}

When studying the representation theory of a group $G$, it is useful to introduce the algebra $ \mC [G ] $ which consists of formal linear combinations of group elements, equipped with the multiplication inherited from the group. In the group algebra $ \mC [G ] $, for each conjugacy class, we can form a sum over all the elements in the conjugacy class of $G$. Such class sums commute with any element of $G$ and form the central sub-algebra of $ \mC [G ]$, i.e. the sub-algebra which commutes with all $ \mC [G ]$. We will refer to $ \cZ[\mC [ G] ] $ as the centre of $ \mC [G ]$. 
Conjugacy classes are in 1-1 correspondence with irreducible representations and there is a basis of the centre consisting of projectors of the form 
\bea 
P_{ R } = { d_R \over |G| } \sum_{ g \in G } \chi_R ( g ) ~ g^{-1} 
\eea
Of primary interest to us is the group algebra of $\mC [ S_n] $ and its centre $ \cZ [ \mC [ S_n] ]$. 
The elements in $\cZ [ \mC [ S_n] ]$ are sums over conjugacy classes $t$ of $ S_n$
\begin{align} 
T = \sum_{ \sigma \in t } \sigma 
\end{align}
Given any $ \sigma \in S_n$, we can generate an element of this subalgebra by summing 
over $ \gamma \in H$. 
\begin{align}\label{group averaging}
\sum_{ \gamma \in S_n } \gamma \sigma \gamma^{-1} 
\end{align}
Some properties of group algebras and their centre can be found in \cite{Hamermesh,goodman2000representations}. 
In the context of AdS/CFT , group algebras $\mC [ S_n] $ and associated representation theory play a role in the half-BPS sector of $\N=4$ SYM in 4D \cite{CJR01,CR02} and also in the symmetric orbifolds in AdS3/CFT2 \cite{MS98,LM99}. Motivated by developments in AdS/CFT we will introduce a generalization of this construction.

\noindent 
{\bf Definition:} Consider an associative algebra $ \mA $ containing a sub-algebra $ \mB= \mC [ H] $, the group algebra of a finite group $H$. Now define the sub-space of $ \mA $ of elements which are invariant under conjugation by $H$. This subspace will contain group averages of the form
\begin{align}
\sum_{ \gamma \in H } \gamma \sigma \gamma^{-1} \,,\qquad \sigma\in \mA
\end{align}
which commute with elements of $\mB$. 
It is easy to verify that these sub-spaces are sub-algebras. We have
\begin{align}
\left(\sum_{ \gamma_1 \in H } \gamma_1 \sigma \gamma_1^{-1}\right)\,
\left(\sum_{ \gamma_2 \in H } \gamma_2 \sigma \gamma_2^{-1}\right)=
\sum_{\gamma_1 \in H} \gamma_1 \left(\sum_{\gamma_3\in H}\sigma_1\gamma_3 \sigma_2\gamma_3^{-1}\right)\gamma_1^{-1}
\end{align}
where we set $\gamma_3 =\gamma_1^{-1}\gamma_2 $. This shows that the product of two group averages is still a group average.
This sub-algebra of $ \mA $ commuting with $ \mB $, in cases where $H$ is a permutation group, will be called a {\it permutation centralizer algebra}. 

Three cases of primary interest will be 
\begin{itemize} 
\item { \bf Example 1 } The algebra $ \mA = \mC [S_n ] $. 
The algebra $ \mB = \mC [S_n ]$. The centralizer of $ \mathbb B $ is $ \Z [ \mC [S_n ] ] $ .

\item {\bf Example 2 } $ \mA = \mC [ S_{m+n}] $ ; $ \mB = \mC [ S_m \times S_n ] $. We will call this algebra $ \cA ( m , n ) $. 

\item { \bf Example 3 } $ \mA = B_N ( m , n ) $ - the walled Brauer algebra ; $ \mB = \mC [ S_m \times S_n ]$. This algebra is called $ \cB_N ( m , n ) $. 

\item { \bf Example 4} $ \mA = \mC [ S_{n} \times S_n ] $ ; $ \mB = \mC [S_n ] $ where the latter is the $S_n$ diagonally embedded in the product group. 
This should be called $ \cK ( n ) $. 

\end{itemize} 
The case where $\mA$ is itself a group algebra has been studied in mathematics, for example, in \cite{DEM2013}.

Our primary interest in this paper will be in $ \cA ( m , n ) $ of example $2$. 
$ \Z [ \mC [S_n ] ] $ of Example 1 will be a useful guide and a source of analogies in our investigations. 
Fourier transformation on $ \cA ( m , n )$ will be related to restricted Schur operators studied in AdS/CFT. 
These are parametrised by representation theory data $ ( R , R_1 , R_2 , i , j ) $ consisting of Young diagrams $ R_1 , R_2 , R $ with $ m , n , m+n $ boxes as well as multiplicity indices $ i , j $. The latter take values $ 1 \le i ,j \le g ( R_1 , R_2 ; R ) $ where $ g ( R_1 , R_2 ; R )$ is the LR multiplicity for the triple of Young diagrams computed with the LR combinatoric rule (see for example \cite{FulHar}). Unlike $ \Z [ \mC [S_n ] ]$, the algebra $ \cA ( m , n )$ is not commutative. 
The central sub-algebra $ \cZ ( m ,n ) $, consisting of the subspace $ \cZ ( m , n ) \subset \cA ( m ,n ) $ which commutes with all of $ \cA ( m ,n ) $ will play a predominant role. Likewise the algebras $ \cB_N ( m , n ) $ and $ \cK ( n ) $ in Examples 3 and 4 are non-commutative.

\vskip.5cm 

\section{Structure of the $ \cA(m ,n)$ algebra }\label{sec:struct} 

The algebra $\A(m,n)$ is constructed by taking all the elements in $\mathbb C[S_{m+n}]$ which are invariant under $\mathbb C[S_m\times S_n]$. Any element of $\sigma \in \mathbb C[S_{m+n}]$ can be mapped to a $\bar\sigma\in \A(m,n)$ by the group averaging
\begin{align}
\bar\sigma =\sum_{\gamma\in S_m\times S_n}\gamma^{-1}\sigma\gamma
\end{align}
The $\bar\sigma$ are formal sums of permutations $\tau$ lying in the same orbit of $\sigma$ under the $\,S_m\times S_n$ action. Each $\tau$ has a stabiliser group, given by those $\gamma\in S_m\times S_n$ for which
\begin{align}
\gamma^{-1}\tau \gamma = \tau
\end{align}
The stabilisers of two permutations $ \tau_1 , \tau_2$  in the same orbit  are generally different (they are conjugate to each other), but they have the same dimension. By the Orbit-Stabiliser theorem, $\bar\sigma$ is then a sum of permutations weighted by the same coefficient:
\begin{align}\label{bar sigma def orbit}
\bar\sigma=|\text{Aut}_{S_m\times S_n}(\sigma)|\sum_{\tau\in \text{Orbit}(\sigma,\,S_m\times S_n)}\tau
\end{align}

$\A(m,n)$ is a finite-dimensional associative algebra (the associativity follows from the associativity of $\mathbb C[S_{m+n}]$), which we can equip with the non-degenerate symmetric bilinear form
\begin{align}\label{A[m,n] bilinear form}
\la\bar \sigma_1,\bar\sigma_2\ra=
%\sum_{ \gamma \in S_m\times S_n } \delta ( \bar\sigma_1 \gamma \bar\sigma_2 \gamma^{-1} )\,,
\delta ( \bar\sigma_1 \bar\sigma_2 )\,,
\qquad\bar\sigma_{1,2}\in \A(m,n)
\end{align}
Here the delta function on the group algebra $ \mC [ S_{m+n} ] $ is a linear function which obeys $ \delta ( \sigma  ) =1  $ for $ \sigma =1$ 
and $ \delta ( \sigma ) =0$ otherwise. 
 
The non-degeneracy of the bilinear form \eqref{A[m,n] bilinear form} implies that $\A(m,n)$ is semi-simple. According to the Wedderburn-Artin theorem, it can then be decomposed into a direct sum of matrix algebras:
\begin{align}\label{Wedderburn theorem}
\A(m,n)=\bigoplus_{R\vdash m+n\atop R_1\vdash m,\,R_2\vdash n}\text{Span}\{Q^R_{R_1,R_2,i,j};i,j\}
\end{align}
In this equation $R,\,R_1$ and $R_2$ are representations of $S_{m+n}$, $S_m$ and $S_n$ respectively. The integers $i,j$ run over the multiplicity $g(R_1,R_2;R)$ of the branching $R\rightarrow R_1\otimes R_2$: $0\leq i,j\leq g(R_1,R_2;R)$. 
%The numbers $g(R_1,R_2;R)$ are called Littlewood-Richardson (LR) coefficients, and they are briefly reviewed in Appendix \ref{Appendix: LR rule for hooks}. 
An explicit expression for $Q^R_{R_1,R_2,i,j}$ is given in terms of the restricted Schur characters \cite{BCK08,KR2,quivcalc}, defined as 
\begin{align}\label{restricted schur def}
\chi^R_{R_1,R_2,i,j}(\sigma) = D^R_{m,m'}(\sigma)\,\,
B^{R\rightarrow R_1,R_2;i}_{m'\rightarrow l_1,l_2}\,\,
B^{R\rightarrow R_1,R_2;j}_{m\rightarrow l_1,l_2}
\end{align}
Here $D^R_{m,m'}(\sigma)$ are the matrix elements of $\sigma$ in the irreducible representation $R$. $B^{R\rightarrow R_1,R_2;j}_{m\rightarrow l_1,l_2}$ is the branching coefficient for the representation branching $R\rightarrow R_1\otimes R_2$, in the $j$-th copy of $R_1\otimes R_2\subset R$. $l_{1,2}$ are states in $R_{1,2}$. 
The restricted Schur characters $\chi^R_{R_1,R_2,i,j}(\sigma)$ are invariant under conjugation by $\mathbb C[S_m\times S_n]$ elements.
With these definitions we can write 
\begin{align}\label{basis perm form}
Q^R_{R_1,R_2,i,j}=\sum_\sigma\chi^R_{R_1,R_2,i,j}(\sigma)\,\sigma
\end{align}
which is manifestly invariant under the action of $\mathbb C[S_m\times S_n]$.
It follows that 
\begin{align}\label{Qprods}
Q^R_{ R_1 , R_2 , i , j } Q^{ S}_{ S_1 , S_2 , k , l } 
=\delta^{R,S}\,\delta_{ R_1 , S_1 }\,\delta_{ R_2 , S_2 } (\delta_{ j k }\,Q^R_{ R_1 , R_2 , i , l }) 
\end{align}
This is in accordance with the decomposition \eqref{Wedderburn theorem}. 
Consequently it is useful to write $Q^R_{R_1,R_2,i,j}$ as
\begin{align}\label{basis ketbra form}
Q^{R}_{R_1,R_2,i,j}=\sum_{m_1,m_2}|R\rightarrow R_1,R_2,\, m_1,m_2,i\ra\la R\rightarrow R_1,R_2,\, m_1,m_2,j|
\end{align}

Moreover, the basis $\left\{Q^{R}_{R_1,R_2,i,j}\right\}$ is complete as we now explain. 
 The number of distinct $Q^{R}_{R_1,R_2,i,j}$'s is equal to the number of restricted Schur characters, which is in turn equal to $ \sum_{ R_1 , R_2 , R } g(R_1,R_2;R)^2$.
 On the other hand the dimension of $\A(m,n)$ is by definition equal to the number of elements of $\mathbb C[S_{m+n}]$ invariant under the $\mathbb C[S_m\times S_n]$ action. Using the Burnside lemma, it is possible to show that this dimension $|\A(m,n)|$ is given as 
\begin{align} 
|\A(m,n)|=\sum_{ R_1 \vdash m , R_2 \vdash n\atop R\vdash m+n } g ( R_1 , R_2 , R )^2 
\end{align}

In each of the blocks in \eqref{Wedderburn theorem} there is a projector of the form $P^R_{R_1,R_2}=\sum_i Q^{R}_{R_1,R_2,i,i}$. Let now $P_R$, $P_{R_1}$ and $P_{R_2}$ be the projectors onto the irreps $R, R_1$ and $R_2$ of $S_{m+n}$, $S_m$ and $S_n$ respectively. Since
\begin{align} 
& \lara{R \rightarrow R_1 , R_2 , m_1 , m_2 , i | P_R P_{ R_1} P_{ R_2} | R \rightarrow R_1 , R_2 , m_1' , m_2' , j }\\[3mm]
&=
\lara{R \rightarrow R_1 , R_2 , m_1 , m_2 , i | P^R_{ R_1,R_2} | R \rightarrow R_1 , R_2 , m_1' , m_2' , j }
= \delta_{m_1 , m_1'} \delta_{ m_2 , m_2'} \delta_{ i , j }\nonumber
\end{align}
for all triplets $R,\,R_1,\,R_2$, we can write
\begin{align}\label{PP}
P^R_{ R_1,R_2} = P_R P_{ R_1} P_{ R_2}
\end{align}
so that the projectors $P^R_{ R_1,R_2}$ are just products of ordinary $S_{m+n}$, $S_{m}$ and $S_{n}$ projectors.
The set $\{P^R_{R_1,R_2}\}$ forms a basis for the centre of $\A(m,n)$, which we call $\Z(m,n)$. Its dimension is then given by the number of non vanishing LR coefficients $g(R_1,R_2;R)$, or
\begin{align}
|\Z(m,n)|=\sum_{ R_1 \vdash m , R_2 \vdash n\atop R\vdash m+n } (1-\delta(g(R_1,R_2;R)))
\end{align}
Here $\delta(g(R_1,R_2;R))=1$ if $g(R_1,R_2;R)=0$ and $\delta(g(R_1,R_2;R))=0$ otherwise.
The generating function for the dimension of the centre is \cite{Bianchi:2006ti}
\begin{align} 
\Z(x,y)=\prod_i { 1 \over ( 1 - x^i - y^i ) } 
\end{align}

We will now argue that the collection of the generators of the centres of $\mathbb C[S_{m+n}]$, $\mathbb C[S_{m}]$ and $\mathbb C[S_{n}]$, that we denote as $\{T_{p}^{(m+n)}\}$, $\{T_{q_1}^{(m)}\}$ and $\{T_{q_2}^{(n)}\}$ respectively, is a set of generators for $\Z(m,n)$. Here $p$, $q_1$ and $q_2$ are integer partitions of $m+n$, $m$ and $n$ respectively. For example, for the partition $ p=( p_1 , p_2 , ... ) $ of $m+n$, the operator $T_{ p}^{(m+n)}$ consists of a sum over permutations belonging to the conjugacy class $p=( p_1 , p_2 , ... )$:
\begin{align}
T_{ p}^{(m+n)} =
 \sum_{ i_1 , \cdots , i_{ p_1 +p_2 + \cdots } 
 \in [ m + n ] } ( i_1 i_2 \cdots i_{p_1} ) 
( i_{ p_1 +1 } i_{ p_1 + 2 } \cdots i_{ p_1 + p_2 } ) \cdots 
\end{align}
$T_{ p}^{(m+n)}$ are sums of conjugates by elements of $S_{m+n}$, whereas $T_{ q_1}^{(m)}$ and $T_{ q_2}^{(n)}$ are sums over $S_{m}\subset S_{m+n}$ and $S_{n}\subset S_{m+n}$ respectively.
To show that $\{T_{p}^{(m+n)},\,T_{q_1}^{(m)},\,T_{q_2}^{(n)}\}$ generate the whole centre $\Z(m,n)$ we can use the following argument. Using the Wedderburn-Artin decomposition \eqref{Wedderburn theorem}, we see that the centre of $\A(m,n)$ is the direct sum of the centres of the matrix algebras $\text{Span}\{Q^R_{R_1,R_2,i,j};i,j\}$. For each of these matrix blocks, that is for any fixed representations $R,\,R_1,\,R_2$ for which $g(R_1,R_2;R)\neq 0$, the centre is one-dimensional, and is spanned by
\begin{align}
P^R_{R_1,R_2}=\sum_{i=1}Q^R_{R_1,R_2,i,i}
\end{align}
Using the equation \eqref{Qprods}, it is immediate to check that
\begin{align}
[P^R_{R_1,R_2}, Q^R_{R_1,R_2,i,j}] = 0\,,\qquad\quad \forall \,\,i,j
\end{align}
We know that $P^R_{R_1,R_2}=P_RP_{R_1}P_{R_2}$, with $P_R$, $P_{R_1}$ and $P_{R_2}$ projectors on the representations $R$, $R_1$ and $R_2$. Therefore every central element of $\A(m,n)$ can be generated with the collection of projectors $\{P_R,\,P_{R_1},\,P_{R_2}\}$.
 For an  $R$ irrep of $S_n$, the projector is 
\begin{align}\label{projector definition}
P_R=\frac{1}{n!}\sum_{\sigma\in S_n}\chi_R(\sigma)\,\sigma = \frac{1}{n!}\sum_{p \in \text{Partitions}(n)}\chi_R(\sigma_p)\,T_p^{(n)}
\end{align}
where $\sigma_p$ is a representative permutation belonging to the conjugacy class $p\vdash n$. This means that every projector $P_R$ can be written as a linear combination of the central elements $\{T_p^{(n)}\}$. We can then write the set $\{P_R,\,P_{R_1},\,P_{R_2}\}$ in terms of the central elements $\{T_{p}^{(m+n)},\,T_{q_1}^{(m)},\,T_{q_2}^{(n)}\}$. Since we know that the former generates the whole $\Z(m,n)$, we can now conclude that the latter is a complete set of generators for the centre $\Z(m,n)$ as well. The basis thus obtained will be useful in the following sections. However, it is important to point out that such a basis is overcomplete. An easy way to see it is to note that, given \eqref{PP}, 
 $P_R\,P_{R_1}P_{R_2}=0$ if $g(R_1,R_2,R)=0$. Therefore, taking a triplet $(R_1,R_2, R )$ for which $g(R_1,R_2,R)=0$ we have, using \eqref{projector definition}:
\begin{align}
\frac{1}{(m+n)!m!n!}\sum_{p \vdash(m+n)\atop q_1 \vdash m,\,q_2\vdash n }\chi_R(\sigma_p)\chi_{R_1}(\sigma_{q_1})\chi_{R_2}(\sigma_{q_2})\,T_p^{(m+n)}\,T_{q_1}^{(m)}\,T_{q_2}^{(n)}=0
\end{align}
This shows that $\{T_{p}^{(m+n)},\,T_{q_1}^{(m)},\,T_{q_2}^{(n)}\}$ is indeed an overcomplete basis.

We can also argue that $\{T_{p}^{(m+n)},\,T_{q_1}^{(m)},\,T_{q_2}^{(n)}\}$ generate $\Z(m,n)$ just by using the Schur-Weyl duality as in \cite{KR2}.
 The $T^{(m)}$ elements are Schur-Weyl dual to $ U(N)$ Casimirs of acting on the upper $m$ indices of $X$-type matrices. This action is generated by
\begin{align} 
(E_x)^i_j = (D_x)^i_j=X^i_l\frac{\partial}{\partial X^j_l}
\end{align}
The $T^{(n)}$ elements are Schur-Weyl dual to $ U(N)$ Casimirs acting on the upper $n$ indices of $Y$-type matrices. We have
\begin{align} 
(E_y)^i_j = (D_y)^i_j=Y^i_l\frac{\partial}{\partial Y^j_l}
\end{align}
Finally, the $T^{(m+n)}$ elements are Schur-Weyl dual to $ U(N)$ Casimirs acting on the upper $n$ and $m$ indices of both $X$- and $Y$-type matrices, and the generator is
\begin{align} 
E^i_j = (E_x)^i_j+(E_y)^i_j
\end{align}
We then have three distinct types of Casimirs:
\begin{align} 
&C_k^{(m+n)} = E^{ i_1}_{i_2} E^{i_2}_{i_3} \cdots E^{i_k}_{i_1} \nn
&C_k^{(m)} = (E_x)^{ i_1}_{i_2} (E_x)^{i_2}_{i_3} \cdots (E_x)^{i_k}_{i_1} \nn
&C_k^{(n)} = (E_y)^{ i_1}_{i_2} (E_y)^{i_2}_{i_3} \cdots (E_y)^{i_k}_{i_1}
\end{align}
But the $C_k^{(m+n)}$, the $C_k^{(m)}$ and the $C_k^{(n)}$ operators measure respectively the $R$, $R_1$ and $R_2$ labels of the restricted Schurs $\chi^R_{R_1,R_2,i,j}$. Therefore they can be used to isolate every subspace $R_1\otimes R_2\subseteq R$, and to build all the correspondent projectors $P^R_{R_1,R_2}$. Since we know that each of these projectors is in a 1-1 correspondence with an element of $\Z(m,n)$, the whole centre $\Z(m,n)$ is obtained.

On the other hand, non-central elements are needed to measure the multiplicity labels $i,j$. This observation will be developed in section \ref{Section: Star product}.

\subsection{Symmetric Group characters and the pairing on the centre $\Z(m,n)$ } 
A central element $ Z_a \in \Z(m,n)$ can be expanded in terms of the projectors $ P^R_{ R_1 , R_2 } $ as
\begin{align} 
Z_a = \sum_{R,R_1,R_2}Z_a^{ R , R_1 , R_2 } P^R_{ R_1 , R_2 }
\end{align}
We can then define
\begin{align} 
\chi^{ R }_{ R_1 , R_2 ; i , j } ( Z_a ) &=\sum_{m_1,m_2} \langle R \rightarrow R_1 , R_2 ,m_1,m_2, i | Z_a | R \rightarrow R_1 , R_2 , m_1,m_2,j \rangle\nn
&=
\delta_{ i j }\sum_{S,S_1,S_2}\sum_{m_1,m_2} Z_a^{ S , S_1 , S_2 } \langle R \rightarrow R_1 , R_2 ,m_1,m_2, i | P^S_{ S_1 ,S_2 } | R\rightarrow R_1 , R_2 ,m_1,m_2, j \rangle\nn
& = \delta_{ i j } Z_a^{ R , R_1 , R_2 } d_{R_1}d_{R_2}
\end{align}
and
\begin{align} 
\chi^{ R }_{ R_1 , R_2 } ( Z_a )=\sum_{i}\chi^{ R }_{ R_1 , R_2 ; i , i } ( Z_a )
 = Z_a^{ R , R_1 , R_2 } g ( R_1 , R_2 , R ) d_{ R_1} d_{ R_2} 
\end{align}
From these equations it also follows that for any central element $Z_a$
\begin{align}
\chi^R_{R_1,R_2,i,j}(Z_a)=\frac{\delta_{i,j}}{g(R_1,R_2;R)}\,\chi^R_{R_1,R_2}(Z_a)
\end{align}

Another useful expansion is in terms of $\{T_{p}^{(m+n)}\}$, $\{T_{q_1}^{(m)}\}$ and $\{T_{q_2}^{(n)}\}$. %We define here the shorthand notation 
%\begin{align} 
%T_{[\vec q\,]}^{(m,n)}=T_{[q_1]}^{(m)}\,T_{[q_2]}^{(n)} 
%\end{align}
%where $\vec q=(q_1,q_2)$. 
Since these elements generate the centre, we can write
\begin{align}\label{ZapqExp} 
Z_a = Z_a^{p,q_1,q_2}\, T^{ ( m , n ) }_p \,T^{ (m)}_{q_1}\, T^{ (n)}_{q_2}
\end{align}
for some $Z_a^{p,q_1,q_2}$ coefficients. However, since the basis generated by $\{T_{p}^{(m+n)},\,T_{q_1}^{(m)},\,T_{q_2}^{(n)}\}$ is overcomplete, such coefficients are not unique. Using the expansion (\ref{ZapqExp}), we can write 
\begin{align}\label{Char exp Zapq, ij}
\chi^R_{ R_1 , R_2 , i , j } ( Z_a ) = { \delta_{ i j } } \,Z_a^{p,q_1,q_2}\,
{ \chi_R ( T^{ (m+n)}_p ) \over d_R } \chi_{R_1} ( T^{(m)}_{q_1} ) \chi_{R_2} ( T^{(n)}_{ q_2} ) 
\end{align}
and
\begin{align}\label{Char exp Zapq}
\chi^R_{ R_1 , R_2 } ( Z_a ) = \sum_i\chi^R_{ R_1 , R_2,i,i } ( Z_a ) = Z_a^{ p, q_1,q_2 } g ( R_1 , R_2 , R ) { \chi_R ( T^{ (m+n)}_p ) \over d_R } \chi_{R_1} ( T^{(m)}_{q_1} ) \chi_{R_2} ( T^{(n)}_{ q_2} ) 
\end{align}
From these equation we see that all the restricted characters of central elements are determined by characters of $ S_{m+n} , S_m , S_n$. Just as the centre of $ S_n$ is generated by class sums, which are dual to irreducible characters of $S_n$, the centre $ \cZ (m,n)$ of $ \cA(m,n)$ is dual to the characters $ \chi^R_{ R_1 , R_2}$ which are nothing by products of characters. 
Therefore, to compute restricted characters of elements in $\Z(m,n)$ we only need the ordinary symmetric group character theory.

We will now use some of the known equations for the character of symmetric group and use them to compute restricted characters in $\cZ( m , n) $. Our aim will be to compute the dual pairing \eqref{A[m,n] bilinear form} for central elements. Equation (B.12) in \cite{quivcalc} reads
\begin{align} \label{B12 quivcalc eq}
{ ( m+n)! \over m! n! } \sum_{ \gamma \in S_m \times S_n } \delta ( \sigma \gamma \tau \gamma^{-1} ) 
=
\sum_{ R , R_1 , R_2 , i , j } 
{ d_{R} \over d_{ R_1} d_{R_2} } \chi^R_{ R_1 , R_2 , i , j } ( \sigma ) \chi^R_{ R_1 , R_2 , i , j } ( \tau ) 
\end{align}
By setting $ \tau =1$ this equation simplifies to 
\begin{align}\label{delta(sigma) char expansion}
( m+n ) ! \delta ( \sigma ) 
=
\sum_{ R } d_R \chi^{ R }_{ R_1 , R_2 , i , i } ( \sigma ) 
\end{align}
where we used
\begin{align} 
\chi^R_{ R_1 , R_2 , i , j } ( 1 ) = \delta_{ i j } d_{ R_1} d_{ R_2} 
\end{align}
We can immediately use this result to show that $\delta(Q^R_{R_1,R_2,i,j})=\delta_{ i j } d_{ R_1} d_{ R_2} $. This is because, using \eqref{basis perm form}
\begin{align}
\delta\left(Q^R_{R_1,R_2,i,j}\right) = 
\sum_\sigma\chi^R_{R_1,R_2,i,j}\delta\left(\sigma\right)=
\chi^R_{R_1,R_2,i,j}(1) = 
\delta_{ i j } d_{ R_1} d_{ R_2} 
\end{align}
It is also worthwhile to notice that, for $\cO\in \A(m,n)$, $\Tr(\cO)=\delta(\cO)$. Therefore we could have obtained the same result by considering
\begin{align}
\Tr(Q^R_{R_1,R_2,i,j})&=\sum_{S,S_1,S_2}
\,\sum_{m_1,m_2\atop m_1', m_2'}\,\sum_k\left\langle S\rightarrow S_1,S_2,m_1',m_2',k|
R\rightarrow R_1,R_2,\, m_1,m_2,i\ra\right.\nn
&\qquad\qquad\times
\la R\rightarrow R_1,R_2,\, m_1,m_2,j
|S\rightarrow S_1,S_2,m_1',m_2',k\ra\nn
&\qquad\qquad
=\delta_{ i j } d_{ R_1} d_{ R_2}
\end{align}
where we used the definition \eqref{basis ketbra form}.

Let us now go back to eq. \eqref{delta(sigma) char expansion}. If we replace $ \sigma $ by a central element $Z_a$, using the expansion \eqref{ZapqExp} and eq. \eqref{Char exp Zapq}, we find
\begin{align} 
( m +n )!\,\delta ( Z_a ) 
=
\sum_{ R , R_1 , R_2 } Z_a^{ p , q_1,q_2 } g ( R_1 , R_2 , R ) 
{ \chi_R ( T^{ (m+n)}_p ) } \chi_{R_1} ( T^{(m)}_{q_1} ) \chi_{R_2} ( T^{(n)}_{ q_2} )
\end{align}
By further replacing $ \sigma \rightarrow Z_a , \tau \rightarrow Z_b $ in \eqref{B12 quivcalc eq} we get, in a similar fashion
\begin{align}\label{centralpairing}
( m+n)!\,& \delta ( Z_a Z_b ) = \sum_{ R , R_1 , R_2,i,j } \frac{ d_R}{ d_{ R_1} d_{ R_2}} \chi^R_{ R_1 , R_2,i,j } ( Z_a ) \chi^R_{ R_1 , R_2,i,j } ( Z_b )\nn
& = Z_a^{ p, q_1,q_2 } Z_{b}^{ p', q_1',q_2' } \sum_{ R , R_1 , R_2 } \frac{g ( R_1 , R_2 , R )}{ d_{ R } d_{ R_1} d_{ R_2 }} \times\\[3mm]
&\qquad\quad\times
\chi_R ( T_p^{(m+n)} )\, \chi_{ R_1 } ( T_{q_1}^{ ( m ) } )\, \chi_{R_2}( T_{q_2}^{ ( n ) } )
\chi_R ( T_{p'}^{(m+n)})\, \chi_{ R_1 } ( T_{q_1'}^{ ( m ) })\, \chi_{R_2}( T_{q_2'}^{ ( n ) } ) \nonumber
\end{align}
Comparing the LHS above with eq. \eqref{A[m,n] bilinear form} %, and recalling that $\chi^R_{R_1,R_2,i,j}(\sigma)=\chi^R_{R_1,R_2,i,j}(\sigma^{-1})$ \td{works also for algebra elements?}, 
we find that for central elements $Z_a,\,Z_b$
\begin{align}
\left\la Z_a,Z_b\right\ra= &\,Z_a^{ p, q_1,q_2 }\, Z_{b}^{ p', q_1',q_2' }\,\frac{1}{(m+n)!} \sum_{ R , R_1 , R_2 } \frac{g ( R_1 , R_2 , R )}{ d_{ R } d_{ R_1} d_{ R_2 }} \times\\[3mm]
&\qquad\quad\times
\chi_R ( T_p^{(m+n)} )\, \chi_{ R_1 } ( T_{q_1}^{ ( m ) } )\, \chi_{R_2}( T_{q_2}^{ ( n ) } )
\chi_R ( T_{p'}^{(m+n)})\, \chi_{ R_1 } ( T_{q_1'}^{ ( m ) })\, \chi_{R_2}( T_{q_2'}^{ ( n ) } ) \nonumber
\end{align}
Thus we have an explicit way of computing the dual paring on the centre $\Z(m,n)$ in terms of ordinary $S_n$ characters.

Similarly, there is a character expansion for $\delta ( Z_a Z_b Z_c )$. We begin by writing
\begin{align}\label{ZZZ delta}
( m+n)!\,& \delta ( Z_a Z_bZ_c ) = \sum_{ R , R_1 , R_2,i,j } \frac{ d_R}{ d_{ R_1} d_{ R_2}} \chi^R_{ R_1 , R_2,i,j } ( Z_aZ_b ) \chi^R_{ R_1 , R_2,i,j } ( Z_c )\nn
&=
\sum_{ R , R_1 , R_2 } \frac{ d_R}{ d_{ R_1} d_{ R_2}g(R_1,R_2;R)} \chi^R_{ R_1 , R_2} ( Z_aZ_b ) \chi^R_{ R_1 , R_2 } ( Z_c )
\end{align}
Since $Z_a$ is central, $Z_a=(c_a)^R_{R_1,R_2}1$, where $(c_a)^R_{R_1,R_2}$ is a constant. This constant can be obtained by considering:
\begin{align}
\chi^R_{ R_1 , R_2,i,j } ( Z_a )=(c_a)^R_{R_1,R_2}\chi^R_{ R_1 , R_2 } ( 1)=(c_a)^R_{R_1,R_2}\,d_{R_1}\,d_{R_2}g(R_1,R_2;R)
\end{align}
We therefore have that
\begin{align}\label{restricted Schur factorisation for Zs}
\chi^R_{ R_1 , R_2 } ( Z_aZ_b ) =\frac{\chi^R_{ R_1 , R_2 } ( Z_a )\,\chi^R_{ R_1 , R_2} ( Z_b )}{d_{R_1}\,d_{R_2}\,g(R_1,R_2;R)}
\end{align}
Using \eqref{restricted Schur factorisation for Zs} in \eqref{ZZZ delta}, and then exploiting \eqref{Char exp Zapq}, we obtain
\begin{align}\label{ZZZ delta full}
&( m+n)!\, \delta ( Z_a Z_bZ_c ) = \sum_{ R , R_1 , R_2 } \frac{ d_R}{ d_{ R_1}^2 d_{ R_2}^2\,g(R_1,R_2;R)^2}\, \chi^R_{ R_1 , R_2 } ( Z_a)\chi^R_{ R_1 , R_2 }(Z_b ) \chi^R_{ R_1 , R_2 } ( Z_c )\nn
&=
Z_a^{p,q_1,q_2}\,Z_b^{p',q_1',q_2'}\,Z_c^{p'',q_1'',q_2''}\,\sum_{ R , R_1 , R_2}\, \frac{g(R_1,R_2;R)}{ d_R^2\,d_{ R_1}^2\, d_{ R_2}^2}\,\chi_R ( T_p^{(m+n)} )\, \chi_{ R_1 } ( T_{q_1}^{ ( m ) } )\, \chi_{R_2}( T_{q_2}^{ ( n ) } )\times\nn
&\qquad\times
\chi_R ( T_{p'}^{(m+n)})\, \chi_{ R_1 } ( T_{q_1'}^{ ( m ) })\, \chi_{R_2}( T_{q_2'}^{ ( n ) } )\,
\chi_R ( T_{p''}^{(m+n)})\, \chi_{ R_1 } ( T_{q_1''}^{ ( m ) } )\, \chi_{R_2}( T_{q_2''}^{ ( n ) } )
\end{align}
More generally, we can use \eqref{restricted Schur factorisation for Zs} to compute the identity coefficient of an arbitrary large products of central elements, $\delta(Z_aZ_b\cdots Z_k)$, just by using ordinary symmetric group characters.

\subsection{Maximal commuting subalgebra}\label{Maximal commuting subalgebra}

In this section we describe the Maximal commuting subalgebra $\cM(m,n)$ of $\A(m,n)$:
\begin{align}
\Z(m,n)\subseteq \cM(m,n)\subseteq\A(m,n)
\end{align}
We often refer to $\cM(m,n)$ as the Cartan subalgebra of $\A(m,n)$. $\cM(m,n)$ is spanned by elements of the form $Q^{R}_{R_1,R_2,i,i}$ (no sum over $i$). For fixed $R_1,\,R_2$ and $R$, the total number of basis elements is $g(R_1,R_2;R)$, so that its dimension is 
\begin{align}
|\cM(m,n)|=\sum_{ R_1 \vdash m , R_2 \vdash n\atop R\vdash m+n } g(R_1,R_2;R)
\end{align}
In Appendix \ref{sec:AnalDim} we derived the dimension formula
\begin{align}\label{dimension formula M}
|\cM ( m ,n ) | %&= \sum_{ p \vdash m } \sum_{ q \vdash n } \cF_{ p_1 , p_2 \cdots } \cF_{ q_1 , q_2 , \cdots } \cF_{ p_1 + p_2 , q_1 + q_2 , \cdots } \prod_{ i } i^{ p_i + q_i } ( p_i + q_i ) ! \cr 
&= \sum_{ p \vdash m } \sum_{ q \vdash n } \cF_p \cF_q \cF_{ p+q }\, Sym ( p+q ) 
\end{align}
where $p,\,q$ are partitions of $m$ and $n$, $\cF_p,\,\cF_q,\,\cF_{p+q}$ are combinatorial quantities dependent only on the partitions $p,\,q$ and $p+q$ respectively, and $ Sym ( p+q )=\prod_{ i } i^{ p_i + q_i } ( p_i + q_i ) ! $ is a symmetry factor.

We now turn to the problem of constructing a basis for $\cM(m,n)$. According to the definition \eqref{basis ketbra form}, to write the basis elements $Q^{R}_{R_1,R_2,i,i}$ we first need to compute the branching coefficients for the branching $R\rightarrow R_1\otimes R_2$. These quantities are in general computationally hard to obtain \footnote{see for example 
a discussion of the difficulty and the simplifications in a ``distant corners approximation'' in \cite{KDGM11}}, and require a choice of a basis in $S_{m+n} $ 
representations adapted to $S_m \times S_n$. However, using the correspondence with matrix algebras given by the Wedderburn-Artin decomposition, 
we can construct the Cartan by solving, in each block, the following equations for $ (g(R_1, R_2 ; R ) -1 ) $ linearly independent elements $Q^R_{R_1,R_2,a} \in \cA ( m , n ) $
\begin{subequations}
\begin{align}
&P^R_{R_1,R_2}\,Q^R_{R_1,R_2,a} = Q^R_{R_1,R_2,a}\\[3mm]
&\lara{P^R_{R_1,R_2},Q^R_{R_1,R_2,a}} = 0\\[3mm]
&\left[Q^R_{R_1,R_2,a} , Q^R_{R_1,R_2,b}\right] = 0
\end{align}
\end{subequations}
In the second equation, we are using the pairing defined in \eqref{A[m,n] bilinear form}. 

\section{Star product for composite operators}\label{Section: Star product} 

In the previous sections we discussed the algebra $\A(m,n)$ and its centre $\Z(m,n)$. We noted that central elements are special, as all their properties only depend on ordinary symmetric group character theory. An example of this is eq. \eqref{ZZZ delta full}. In this section we will take advantage of this fact to compute physically relevant quantities, in particular two and three point functions of BPS operators in $\N=4$ SYM. To do so, we will first start by discussing the one matrix sector  in $\N=4$ SYM, reviewing the permutation description of $U(N)$ matrix invariants which are Gauge Invariants Operators (GIOs) in the conformal field theory. We will stress that for this case there is an underlying $\cZ [ \mathbb{C}[ S_n] ] $ algebra. The one matrix problem will be used as a guide to extend to the two matrix problem, that we treat in subsection \ref{Two matrix problem}. Here the underlying algebra will be $\A(m,n)$.

\subsection{One matrix problem}
Let us consider a matrix invariant constructed with $n$ copies of the same matrix $Z$. Any such invariant can be written in terms of a contraction 
\begin{align}\label{Single matrix op}
\cO_{ \sigma } ( Z ) = tr \left(Z^{ \otimes n } \sigma \right)\,,\qquad\qquad\sigma\in S_n
\end{align}
subject to the equivalence relation
\begin{align}
\cO_{\sigma}(Z)=\cO_{\gamma^{-1}\sigma\gamma}(Z)\,,\qquad\qquad\gamma\in S_n
\end{align}

Polynomials in $Z$ like the one in \eqref{Single matrix op} can be multiplied together. Set $\sigma_1 \in S_{n_1}$, $\sigma_2\in S_{n_2}$. By multiplying together $\cO_{\sigma_1}(Z)$ and $\cO_{\sigma_2}(Z)$ we get
\begin{align}\label{Circ product single matrix}
\cO_{\sigma_1}(Z)\,\cO_{\sigma_2}(Z)=\cO_{\sigma_1 \circ \sigma_2}(Z)
\end{align}
where $\sigma_1 \circ \sigma_2\in S_{n_1}\times S_{n_2} \subset S_{ n_1 + n_2 } $. Therefore for the usual product of matrix invariants,
 $\sigma_1 \circ \sigma_2$ lives in the symmetric group of degree $n_1 + n_2 $. We can define
\begin{align}\label{CS infty}
\mathbb C[S_\infty]=\bigoplus_{n}\mathbb C[S_n]
\end{align}
which is closed under the circle product 
\begin{align}
\circ : \mathbb C[S_\infty]\otimes \mathbb C[S_\infty]\rightarrow \mathbb C[S_\infty]
\end{align}
However, we can define another associative product, that we call \emph{star product}, which closes on the operators of fixed degree:
\begin{align}\label{Star product one matrix}
\cO_{\sigma_1}(Z)*\cO_{\sigma_2}(Z)= \cO_{\sigma_1\sigma_2}(Z)\,,\quad\qquad\sigma_{1,2}\in S_n
\end{align}
It is immediate to see how this product is different from the ordinary GIO multiplication product \eqref{Circ product single matrix}: $\sigma_1,\,\sigma_2$ and $\sigma_1 \sigma_2 $ are all permutations of $n$ elements, and the star product is generally non-commutative.
%\begin{align}\label{Star product one matrix}
%*:\mathbb{C}S_n\times \mathbb C[S_n]\rightarrow \mathbb{C}S_n
%\end{align}
Let $[\sigma]$ be the conjugacy class of $ \sigma $. We now define a map from the multi-trace GIOs to the class-algebra
\begin{align} 
\cO_{ \sigma } ( Z ) \rightarrow { 1 \over \hbox{size of $[\sigma]$ } } \sum_{ \tau \in [\sigma] } \tau \equiv { T_{ \sigma } \over |T_{ \sigma} | } 
\end{align}
This map is 1-1 at large $N$. Let us focus on this case. We can expand the product of $T_i,\, T_j\in \cZ [ \mathbb{C } [S_n] ] $ as
\begin{align} 
T_i T_j = C_{ij}^k T_k 
\end{align}
Here the $C_{ij}^k$ are the class algebra structure constants. By multiplying both sides above by $T_l$ and taking the coefficient of the identity we get
\begin{align}
\delta\left(T_i T_jT_l\right) = C_{ij}^k\, \delta\left(T_k T_l\right)=\delta_{k,l}C_{ij}^k \left|T_l\right| = C_{ij}^k |T_k| 
\end{align}
Now we expand the star product $\cO_{ \sigma_1 } ( Z ) * \cO_{ \sigma_2} ( Z )$ as
\begin{align}\label{star prod}
\cO_{ \sigma_1 } ( Z ) * \cO_{ \sigma_2} ( Z ) =\sum_{ p } { |T_{\sigma_{p}}| \over |T_{\sigma_1}| ~ | T_{\sigma_2}| } C_{ [\sigma_1] [\sigma_2]}^{p} \cO_{\sigma_{ p} } ( Z ) = \sum_{p} { \delta ( T_{ \sigma_1} T_{ \sigma_2} T_{ \sigma_p}) \over |T_{\sigma_1}| ~ | T_{\sigma_2}| } \cO_{ \sigma_{p} } ( Z ) 
\end{align}
where the sum is over the conjugacy classes $p$ of $S_n$. $\sigma_{p}$ is a representative element of the conjugacy class $p$.
This equation will lead to a new expression for the two point functions of GIOs built from $ Z , Z^{\dagger}$ 
in $ \cN =4$ SYM. First observe that setting $ Z $ to the identity $N \times N $ matrix 
\bea 
\cO_{ \sigma } ( Z = 1_N ) = N^{ C_{ \sigma } } 
\eea
where $ C_{ \sigma }$ is the number of cycles in the permutation $ \sigma $. 
Now consider taking the star product of $\cO_{ \sigma_1 } ( Z ) ,\, \cO_{ \sigma_2} ( Z )$ and then setting $Z=1_N$. 
We have, according to \eqref{star prod} 
\begin{align}
&\left.\cO_{ \sigma_1 } ( Z ) * \cO_{ \sigma_2} ( Z )\vphantom{\sum}\right|_{Z=1_N} =\frac{1}{|T_{\sigma_1}|\,| T_{\sigma_2}| }\sum_{p }
\delta ( T_{\sigma_1}T_{\sigma_2}T_{\sigma_p} )\, \cO_{\sigma_p}(1_N)\nn
&\qquad
=\frac{1}{|T_{\sigma_1}|\,| T_{\sigma_2}| }\sum_{ p}
\delta ( T_{\sigma_1}T_{\sigma_2}T_{\sigma_p} )\, N^{C_{\sigma_p}}=
\frac{1}{n! |T_{\sigma_1}|\,| T_{\sigma_2}| }\sum_{\gamma\in S_n}
\delta ( \gamma T_{\sigma_1}\gamma^{-1}T_{\sigma_2}\Omega )
\end{align}
where we set $\Omega=\sum_{p }T_{\sigma_p} N^{C_{\sigma_p}}$. On the other hand the free field correlator is known to be \cite{CJR01}
\begin{align}
\left\la \cO_{\sigma_1} ( Z ) \cO_{\sigma_2}^\dagger ( Z ) \right\ra = { 1 \over |T_{\sigma_1}|\,|T_{\sigma_2}| } \sum_{\gamma\in S_n}
\delta ( \gamma T_{\sigma_1}\gamma^{-1}T_{\sigma_2}\Omega )
\end{align}
so that 
\begin{align}\label{2pt function 1matrix}
\boxed{
\left\la \cO_{\sigma_1}( Z ) \cO_{\sigma_2}^\dagger( Z ) \right\ra=n!\,\left.\cO_{ \sigma_1 } ( Z ) * \cO_{ \sigma_2} ( Z )\vphantom{\sum}\right|_{Z=1_N} 
} 
\end{align}
The two point function $\left\la \cO_{\sigma_1} ( Z )\cO_{\sigma_2}^\dagger ( Z ) \right\ra$ is therefore proportional to the star product $\cO_{ \sigma_1 } ( Z ) * \cO_{ \sigma_2} ( Z )$ followed by the evaluation $Z\rightarrow 1_N$.

Similar considerations lead to the following expression for the extremal three point function. In this case, we find that $\left\la \cO_{\sigma_1}( Z ) \cO_{\sigma_2}( Z ) \cO_{\sigma_3}^\dagger ( Z ) \right\ra$ is proportional to the usual product $\cO_{\sigma_1}( Z ) \,\cO_{\sigma_2}( Z ) $, followed by 
the star product with $\cO_{\sigma_3}( Z ) $, followed by the evaluation $Z\rightarrow 1_N$. To see this, take $\sigma_1\in S_{n_1}$, $\sigma_2\in S_{n_2} $ and consider
\begin{align}
\left.\left(\cO_{\sigma_1}( Z ) \,\cO_{\sigma_2}( Z ) \right)*\cO_{\sigma_3}( Z ) \vphantom{\sum}\right|_{Z=1_N}&=\frac{1}{|T_{\sigma_1\circ\sigma_2}|\,|T_{\sigma_3}|\,}\,\delta\left(T_{\sigma_1\circ\sigma_2}T_{\sigma_3}\Omega\right)
\end{align}
where $T_{\sigma_1\circ\sigma_2}\in \cZ [ \mathbb C[S_{ n_1 + n_2} ] ] $, $T_{\sigma_3}\in \cZ [ \mathbb C[S_{n_1+n_2}] ] $ and $\Omega=\sum_{\sigma\in S_{n_1+ n_2 }}\sigma N^{C_{\sigma}}$.
On the other hand the correlator in $\cN=4$ SYM \cite{CJR01} is 
\begin{align}
\left\la \cO_{\sigma_1}( Z ) \cO_{\sigma_2}( Z ) \cO_{\sigma_3}^\dagger( Z ) \right\ra&=
\sum_{\gamma\in S_{n_1 + n_2} }\delta\left(\gamma(\sigma_1\circ\sigma_2)\gamma^{-1}\sigma_3^{-1}\Omega\right) = 
\frac{ ( n_1 +n_2 )! }{|T_{\sigma_1\circ\sigma_2}|\,|T_{\sigma_3}|} \delta\left(T_{\sigma_1\circ\sigma_2}T_{\sigma_3}\Omega\right) 
\end{align}
 so that
\begin{align}\label{3pt function 1matrix}
\boxed{ 
\left\la \cO_{\sigma_1}( Z ) \cO_{\sigma_2}( Z ) \cO_{\sigma_3}^\dagger( Z ) \right\ra =
(n_1+ n_2) ! \left.\left(\cO_{\sigma_1 \circ \sigma_2}( Z ) \right)*\cO_{\sigma_3}( Z ) \vphantom{\sum}\right|_{Z=1_N}
} 
\end{align}
Given that these correlators are neatly expressed in terms of the star product, it would be interesting to give an interpretation of the latter in the dual $AdS_5\times S_5$ side.

We will now write similar equations for the two matrix problem.

\subsection{Two matrix problem}\label{Two matrix problem}

For the two matrix problem, the GIOs are polynomials in the $X,Y$ matrices. Formally, we can write them in terms of a permutation $\sigma\in S_{m+n}$ as
\begin{align}
\cO_{ \sigma } ( X , Y ) = \Tr\left( X^{ \otimes m } \otimes Y^{ \otimes n }\,\sigma \right) 
\end{align}
As in the one matrix problem, there is an equivalence relation
\begin{align}
\cO_{\sigma}( X , Y ) =\cO_{\gamma \sigma \gamma^{-1}}( X , Y ) \,,\qquad\gamma\in S_m\times S_n
\end{align}

To each of these GIO $\cO_\sigma$ we can associate a specific element $N_{\sigma}$ of $\A(m,n)$ that we call a necklace. We define a necklace $N_\sigma$ as 
\begin{align}
N_\sigma = \frac{1}{|\text{Aut}_{S_m\times S_n}(\sigma)|}\sum_{\gamma\in S_m\times S_n}\gamma\sigma\gamma^{-1}
\end{align}
or equivalently as
\begin{align}
N_{\sigma}=\sum_{\tau\in \text{Orbit}(\sigma,\,S_m\times S_n)}\tau
\end{align}
where the sum is restricted to the permutations $\tau$ in the group orbit of $\sigma$ under $S_m \times S_n$. We can think of the necklaces as the normalised version of the $\bar\sigma$ elements defined in \eqref{bar sigma def orbit}. The set of necklaces form a basis for $\A(m,n)$. We associate a GIO to a necklace simply by mapping
\begin{align}\label{GIOs to necklaces}
\cO_\sigma ( X , Y ) \rightarrow { 1 \over | N_{ \sigma }| } N_{\sigma}
\end{align}
For example, for the GIO corresponding to the permutation $\tilde\sigma=(1,2,4,5)(3,6)\in S_6$:
\begin{align}
\cO_{\tilde\sigma}( X , Y ) =\Tr(X^2Y^2)\Tr(XY)
\end{align}
we associate, through the map \eqref{GIOs to necklaces}, the $\A(3,3)$ element
\begin{align}\label{A(3,3) necklace example}
N_{\tilde \sigma}=\sum_{S_3\times S_3}\gamma\tilde\sigma\gamma^{-1}=\sum_{a_1\neq a_2\neq a_3\in \{1,2,3\}\atop \bar b_1\neq \bar b_2\neq \bar b_3\in \{4,5,6\}}(a_1,a_2,\bar b_1,\bar b_2)(a_3,\bar b_3)
\end{align}
Similarly, for the GIO specified by $\tilde \sigma=(1,2,3)\in S_6$
\begin{align}
\cO_{\tilde\sigma}( X , Y ) = \Tr(X^2Y)\Tr(Y)^3
\end{align}
we associate the $\A(2,4)$ necklace
\begin{align}\label{A(2,4) necklace example}
N_{\tilde \sigma} = \sum_{a_1\neq a_2\in \{1,2\}\atop \bar b_1\in \{3,4,5,6\}}\,(a_1,a_2,\bar b_1)
\end{align}
Notice that in the necklaces we do not explicitly write the single cycle permutations, but rather we leave them implicit. In the last example, these single cycle permutations would account for the multi-trace $\Tr(Y)^3$ component of $\cO_{\tilde\sigma} = \Tr(X^2Y)\Tr(Y)^3$. 

\begin{figure}[H]
\begin{center}\includegraphics[scale=1]{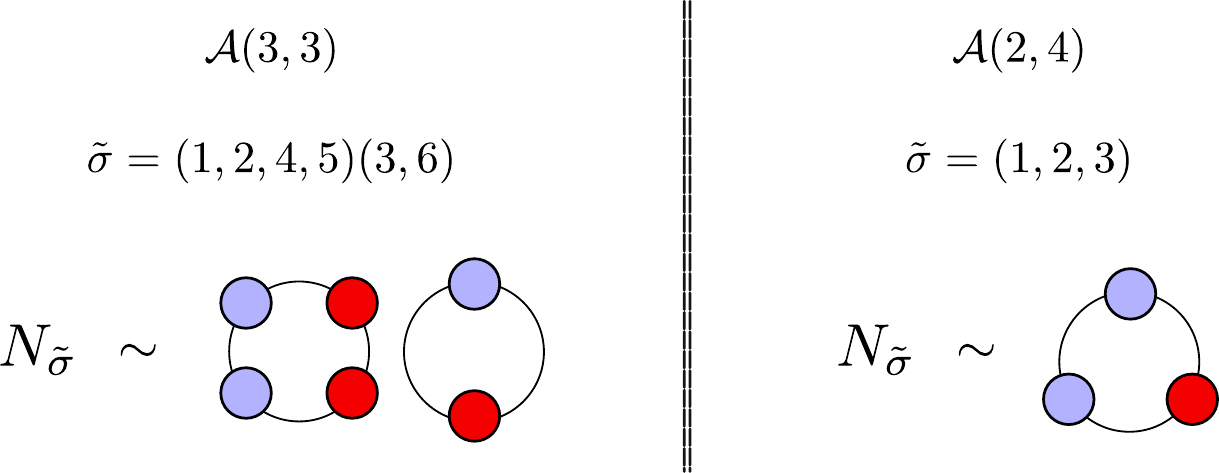}\\[1mm]
\caption{Pictorial interpretations of the necklaces in the examples \eqref{A(3,3) necklace example} and \eqref{A(2,4) necklace example}.}\label{fig: Necklaces example}
\end{center}
\end{figure}

From these examples it is clear how these necklaces are built by taking products of cyclic objects, which in turn are constructed using two different types of \emph{beads}.
Such cyclic objects are well studied in Polya theory. They can be related to 
 the single cycle permutations in $ S_{ m+n}$ with equivalences generated by $ S_m \times S_n$. 
 These equivalence classes form the algebra $\A(m,n)$. 
 We can imagine having blue beads corresponding to integers $[1,2,..m]$ and red beads corresponding to integers $[m+1,m+2,...,m+n]$. Therefore, we can pictorially depict the necklaces of examples \eqref{A(3,3) necklace example} and \eqref{A(2,4) necklace example} as in figure \ref{fig: Necklaces example}. The same structure is present in the GIO $\cO_\sigma$ corresponding to the necklace $N_\sigma$. In this case the single-traces are the cyclic objects, and the role of the blue and red beads is played by the $X$ and $Y$ type fields respectively.

The map \eqref{GIOs to necklaces} is 1-1 at large $N$: as in the 1-matrix problem, we now focus on this case. There is a natural product on the space of two matrix GIOs
coming from multiplying the multi-traces. For such a product, the degrees of the permutations add:
\begin{align}
\cO_{\sigma_1}(X,Y)\cO_{\sigma_2}(X,Y)=\cO_{\sigma_1\circ \sigma_2}(X,Y)
\end{align}
Here $\sigma_1\in S_{m_1 + n_1} $ is a representative of a class in $ \cA ( m_1 , n_1 )$ and 
$\sigma_2\in S_{m_2 + n_2} $ represents a class in $ \cA ( m_2 , n_2 )$, while 
 $\sigma_1\circ \sigma_2 \in S_{m_1 + n_1}\times S_{m_2 + n_2} \subset S_{ m_1 + m_2 + n_1 + n_2 } $ 
represents a class in $ \cA ( m_1 + m_2 , n_1 + n_2 ) $. Continuing the analogy with \eqref{CS infty}, we can define
\begin{align}
\cA ( \infty , \infty ) = \bigoplus_{ m , n } \cA ( m , n ) 
\end{align}
and for $\bar \sigma_1\in\A ( m_1,n_1 ) $ and $\bar \sigma_2\in\A ( m_2,n_2 ) $ we have 
\begin{align}
\circ : \cA ( \infty , \infty ) \otimes \cA ( \infty , \infty ) \rightarrow \cA ( \infty , \infty ) 
\end{align}
As in the one matrix case, there is however a second type of product of GIOs that we can construct. The product on $ \cA ( m ,n ) $ can in fact be used to define a closed and associative star product on the space of the multi-trace operators with fixed numbers $ ( m ,n)$ of $(X, Y)$, in the same fashion as \eqref{Star product one matrix}:
\begin{align}
\cO_{\bar \sigma_1}(X,Y)*\cO_{\bar \sigma_2}(X,Y)=\cO_{\bar \sigma_1 \bar \sigma_2 }(X,Y)\,,\qquad\bar \sigma_{1,2}\in \A(m,n)
\end{align}
Notice that here $\bar \sigma_1,\,\bar \sigma_2$ and $\bar \sigma_1 \bar \sigma_2$ are all of the same degree, and that the star product is non-commutative.
 We will use this star product to express the two point function of GIOs built from $X,\,Y$.

Since the set of necklaces $\{N_a\}$ forms a basis for $\A(m,n)$, we can expand the product $N_aN_b$ as
\begin{align}
N_aN_b=C_{a,b}^cN_c
\end{align}
for some structure constants $C_{a,b}^c$. Moreover, the necklaces are orthogonal in the metric \eqref{A[m,n] bilinear form}:
\begin{align}
\lara{N_a,N_b}=\delta(N_a N_b)=\delta_{a,b}|N_b|
\end{align}
Here $|N_a|$ is the number of permutations in the necklace $N_a$. We can write 
\begin{align}
\delta(N_aN_bN_c) = |N_c|C_{a,b}^c
\end{align}
Now use the map \eqref{GIOs to necklaces} to map the two matrix invariants $\cO_{a}( X , Y ) $ and $\cO_{b}( X , Y ) $ to the necklaces $N_a$ and $N_b$ respectively. Then
\begin{align}
\cO_{a}(X,Y)*\cO_{b}(X,Y)=\sum_c C_{a,b}^c\frac{|N_c|}{|N_a|\,|N_b|}\cO_c(X,Y)=\sum_c\frac{1}{|N_a|\,|N_b|}\,\delta(N_aN_bN_c)\,\cO_c(X,Y)
\end{align}
As for the one matrix problem case, by setting $X=Y=1_N$ we get
\begin{align}
\left.\cO_{a} ( X , Y ) *\cO_{b} ( X , Y ) \vphantom{\sum}\right|_{X=Y=1_N}=\frac{1}{|N_a|\,|N_b|}\,\delta(N_aN_b\Omega)
\end{align}
where $ \Omega = \sum_{ \sigma \in S_{ m+n} } \sigma N^{ C_{ \sigma }} $. 
On the other hand the free field correlator \cite{CR02,BCK08} is 
\begin{align}\label{2pt func 2matrix}
\lara{\cO_a ( X , Y ) \cO_b^\dagger ( X , Y ) }&=\sum_{\gamma\in S_m\times S_n}\delta(\gamma a\gamma^{-1}b^{-1}\,\Omega)=\frac{1}{|N_a|\,|N_b|}\sum_{\gamma\in S_m\times S_n}\delta(\gamma N_a\gamma^{-1}N_b\,\Omega)\nn
&=\frac{m!\,n!}{|N_a|\,|N_b|}\delta( N_aN_b\,\Omega)
\end{align}
Therefore, in analogy with \eqref{2pt function 1matrix} and \eqref{3pt function 1matrix}, we can write the two point function as
\begin{align}\label{{2pt function 2matrix}}
\boxed{ 
\lara{\cO_a ( X , Y ) \cO_b^\dagger ( X , Y ) }=m!\,n!\left.\cO_{a}( X , Y ) *\cO_{b} ( X , Y ) \vphantom{\sum}\right|_{X=Y=1_N}
} 
\end{align}
and the extremal three point function as
\begin{align}\label{3pt function 2matrix}
\lara{\cO_a( X , Y ) \cO_b( X , Y ) \cO_c^\dagger ( X , Y ) }= ( m_1 + m_2) ! ( n_1 + n_2) ! \left.\cO_{a \circ b}( X , Y ) *\cO_c ( X , Y ) \vphantom{\sum}\right|_{X=Y=1_N}
\end{align}
where $a\in S_{m_1+n_1}$, $b\in S_{m_2+n_2}$ and $c\in S_{m_1+n_1+m_2+n_2}$. Finally, notice that the pairing \eqref{A[m,n] bilinear form} is proportional to the planar correlator \cite{Vaman:2002ka,Brown:2010pb,Pasukonis:2010rv} of BPS operators: given $\cO_{a}( X , Y ) $ and $\cO_{b}( X , Y ) $, we have
\begin{align}
\lara{\cO_{a} ( X , Y ) \cO_{b}^\dagger( X , Y )}_{planar}
=
m!n!\lara{a,b}
\end{align}
where the pairing on the RHS is the one in eq. \eqref{A[m,n] bilinear form}. 

Let us now focus on the centre of $\A(m,n)$. In section \ref{sec:struct} we argued that the centre is generated by $\{T_{p}^{(m+n)},\,T_{q_1}^{(m)},\,T_{q_2}^{(n)}\}$. We remind the reader that $\{T_{p}^{(m+n)}\}$, $\{T_{q_1}^{(m)}\}$ and $\{T_{q_2}^{(n)}\}$ are the generators of the centres of $\mathbb C[S_{m+n}]$, $\mathbb C[S_{m}]$ and $\mathbb C[S_{n}]$ respectively, and that $p$, $q_1$ and $q_2$ are integer partitions of $m+n$, $m$ and $n$.
A GIO $\cO_{T_{p}^{(m+n)}}(X,Y)$ can be understood as a descendant of a single matrix 1/2 BPS state $\cO_{T_{p}^{(m+n)}}(X)$ under the $U(2)$ internal symmetry that mixes the $X$ and $Y$ fields. In fact, given $(D^-)^i_j=Y^i_k\frac{\partial}{\partial X^j_k}$: we can write
\begin{align}
\cO_{T_{p}^{(m+n)}}(X,Y) \sim (D^-)^n\cO_{T_{p}^{(m+n)}}(X)
\end{align}
This means that central elements (and their corresponding matrix gauge invariants), described in terms of 
the over-complete basis $\{T_{p}^{(m,n)}\,T_{q_1}^{(m)}\,T_{q_2}^{(n)}\}$, are formed from composites which 
employ both the usual product and the star product :
\begin{align}
[\text{Descendant Operators}]*\{(X \text{-Operators })\,\,(Y \text{-Operators })\}
\end{align}
The descendant GIOs are associated to $T_{p}^{(m+n)}$ elements, $X$- and $Y$- GIOs to $T_{q_1}^{(m)}$ and $T_{q_2}^{(n)}$ elements respectively. 
In terms of the permutations we are taking the product in $ \cA ( m, n ) $ along with the circle product $ \circ : \cA (m,0) \otimes \cA ( 0, n ) \rightarrow \cA ( m , n )$. 

Single-trace symmetrised traces are $U(2)$ descendants of single-trace operators built from a single matrix. In terms of the permutation language, they correspond to single-cycle permutations which are invariant under any reshuffling\footnote{Further details of symmetrised traces in terms of an operation on the permutations in the $ \cO_{ \sigma } ( X , Y ) $ can be found in \cite{Pasukonis:2010rv}.}. 
On the other hand, $U(2)$ descendants of multi-trace operators built from one matrix form a subspace of the space spanned by products of symmetrised single-trace states. In other words, not all products of single-trace descendants are themselves descendants. One way to see this explicitly is the following. Let $ST_{m,n}$ be the space of symmetrised traces with $m$ copies of $X$ and $n$ copies of $Y$ matrices. The generating function  for the dimension $\text{Dim}(ST_{m,n})$ is 
\begin{align}
\prod_{i,j\in \Omega}\frac{1}{1-x^iy^j}=\sum_{m,n}\,\text{Dim}(ST_{m,n})\,x^my^n
\end{align}
where $\Omega = \{0\leq i\leq \infty\}\cup\{0\leq j\leq \infty\}\setminus\{i=j=0\}$.
Let $ST_{m+n}$ be the space of symmetrised traces with a total of $m+n$ matrices, with any number of $X$ or $Y$. 
We have 
\begin{align}
\text{Dim}(ST_{m+n}) = \sum_{i=0}^{m+n}\text{Dim}(ST_{i,m+n-i})
\end{align}
On the other hand, the total number of $U(2)$ descendants obtained from a multi-trace operator with $m+n$ copies of $X$ is 
\begin{align}
(m+n+1)~ p(m+n)
\end{align} 
$p(m+n)$ is the number of partitions of $m+n$ (the number of highest weight states), while $m+n+1$ is the number of descendants for a fixed highest weight. It can now be checked that $\text{Dim}(ST_{m+n})> p(m+n)(m+n+1) $. This indeed proves our original claim.

\subsection{Cartan subalgebra and the minimal set of charges }\label{charges-result} 

In \cite{KR2}, it was observed that, in the free limit, multi-matrix gauge  theories have enhanced symmetries including products of unitary groups. There are Noether charges for these enhanced symmetries. Casimirs constructed from these charges have eigenvalues which can distinguish all the labels $ R , R_1 , R_2 , i , j $ of restricted Schur operators. 
Because of Schur-Weyl duality, these charges are also expressible in terms of permutations. 
Given the definitions in this paper, this action of permutations amounts to the action of 
$ \cA ( m , n )$ on itself by the left or right regular representation. We can now characterize more precisely what is a minimal set of charges which can measure all the labels. In section \ref{Maximal commuting subalgebra} we introduced the Cartan subalgebra $\cM(m,n)$, and gave a prescription to build a basis for it. We need to find a subspace $C_{m,n}$ of $\cM(m,n)$ such that polynomials in some basis elements $c_a \in C_{m,n}$ with coefficients taking values in the centre $\Z(m,n)$ span $\mathcal M(m,n)$. In other words $C_{m,n}$ contains a minimal set of generators for $\cM(m,n)$ as a polynomial algebra over $\Z(m,n)$. A minimal set of generators for $ \Z ( m ,n )$, along with the basis elements of the subspace $C_{m,n}$, provide a complete set of charges, which can measure all the labels of the $ Q^R_{R_1, R_2, i , j }$ by left and right multiplication. Let $N^{min} ( \Z (m,n) ) $ be the minimal number of elements of $ \Z ( m ,n )$ which generate $ \Z ( m , n ) $ as a polynomial algebra. 
Also, let $ N^{min}_{ \Z(m,n) } ( \cM ( m , n ) )$ be the minimal number of elements of $\cM ( m , n )$ which generate $\cM ( m , n )$ as a polynomial algebra over $ \Z ( m , n )$. Left multiplication by these generators correspond to enhanced symmetry charges which 
measure the multiplicity index $i$ of restricted Schur operators. Right multiplication by the same generators correspond to other enhanced symmetry charges which measure the multiplicity index $j$ of restricted Schur operators. Hence the minimal number of charges is 
\bea\label{charge-min} 
\boxed{ 
N^{min} ( \Z (m,n) ) + 2N^{min}_{ \Z(m,n) } ( \cM ( m , n ) )
} 
\eea
An \emph{important open problem} is to determine this function of $(m,n)$ in general. 
This will tell us how many bits of information completely specify all the operators in a multi-matrix set-up. 

The above discussion is complete for the case where $ m+n < N$, which is adequate for a treatment of 
the physics at all orders in the $1/N$ expansion. For finite $N$ effects, where we consider $ m+n > N $, 
the charges given by the above still determine all the multi-matrix invariants, but they are not a minimal set 
any more. The discussion can be easily adapted to this case. 
Define 
\bea 
\cA_N^{ null} ( m , n ) = \bigoplus_{ R \vdash m+n : c_1 ( R ) > N } \bigoplus_{ R_1\vdash m,\,R_2\vdash n}\text{Span}\{Q^R_{R_1,R_2,i,j};i,j\}
\eea
The quotient 
\bea 
 \cA_N ( m , n ) = \cA ( m , n ) / \cA_N^{ null} ( m , n ) 
\eea
is a closed sub-algebra of blocks surviving the finite $N$ cut. 
It has a centre $ \cZ_N ( m ,n ) $ and a Cartan $ \cM_N ( m ,n ) $ which are simply related 
to $ \cZ ( m, n ) $ and $ \cM ( m ,n ) $ by quotienting out the parts belonging to $\cA_N^{ null} ( m , n )$. 
Let $ N^{min} ( \Z_N (m,n) ) $ be the number of generators in a minimal generating set for $\cZ_N ( m ,n ) $ as a polynomial algebra. 
Let $ N^{min}_{ \Z_N (m,n) } ( \cM_N (m,n) ) $ be the number of generators in a minimal generating set for $\cM_N ( m ,n ) $ as a polynomial algebra
over $ \Z_N (m,n)$. The minimal number of charges needed is 
\bea\label{minchargesN}
N^{min} ( \Z_N (m,n) ) + 2N^{min}_{ \Z_N (m,n) } ( \cM_N ( m , n ) )
\eea
We expect \eqref{charge-min},\eqref{minchargesN} will have implications for information theoretic discussions of AdS/CFT such as 
\cite{BBJS05,BCLS06}.

\section{Computation of the finite $N$ correlator}\label{sec:finiteNcorrelator} 

In this section we will derive a finite $N$ generating function for the two point function of operators of the form
\begin{align}\label{ops XmYn}
\cO=\Tr(X^mY^n)
\end{align}
in the free field metric. Operators like the one in \eqref{ops XmYn} correspond to $\A(m,n)$ elements
\begin{align}
\frac{1}{m!n!} T_{\bar 1,1}T_{[m]}^{(X)}T_{[n]}^{(Y)}
\end{align}
where $T_{\bar 1,1}=T_{2}^{(X,Y)}-T_{2}^{(X)}-T_{2}^{(Y)}$. Here $T_{2}^{(X,Y)}$, $T_{2}^{(X)}$ and $T_{2}^{(Y)}$ are the sum of transpositions in $S_{m+n}$, $S_m$ and $S_n$ respectively. $T_{\bar 1,1}$ can be understood as a joining operator, merging the $(1\cdots m)$ type cycles with the $(m+1\cdots m+n)$ type cycles.

The two point function \eqref{2pt func 2matrix} therefore reads, with $\cO=\Tr(X^mY^n)$
\begin{align}\label{OO+}
\la \cO\cO^\dagger\ra&=\frac{1}{m!^2n!^2}\sum_{\gamma\in S_m\times S_n}\,\,\sum_{\sigma\in S_{m+n}}
\delta\left(
\gamma\,\,
 T_{\bar 1,1}T_{[m]}^{(X)}T_{[n]}^{(Y)}\,\,
\gamma^{-1}\,\,
T_{\bar 1,1}T_{[m]}^{(X)}T_{[n]}^{(Y)}\,\,
\sigma
\right)N^{C_{\sigma}}\nn
&=
\frac{1}{m!n!}
%\sum_{\sigma\in S_{m+n}}
\delta\left(
 T_{\bar 1,1}T_{[m]}^{(X)}T_{[n]}^{(Y)}\,\,
T_{\bar 1,1}T_{[m]}^{(X)}T_{[n]}^{(Y)}\,\,
\Omega
\right)
\end{align}
%The $1/m!^2n!^2$ is a normalisation factor. Each $ T_{\bar 1,1}T_{[m]}^{(X)}T_{[n]}^{(Y)}$ consists of a sum $mn*(m-1)!*(n-1)!=m!n!$ permutations all invariant under conjugation of the subgroup $S_m\times S_n$. 
where we set $\Omega=\sum_{\sigma\in S_{m+n}}\sigma N^{C_\sigma}$. This quantity can be computed using only ordinary character theory. Using eq. \eqref{ZZZ delta full} and using the shorthand notation $g=g(R_1,R_2;R)$ we write
\begin{align}\label{chi initial form}
&\la \cO\cO^\dagger\ra=
\frac{1}{(m+n)!m!n!}\,
\sum_{R_1\vdash m\atop R_2\vdash n}\,\sum_{R\vdash m+n}\,
\frac{d_R}{d_{R_1}^2d_{R_2}^2g^2}\,
\left(\chi_{R_1,R_2}^R\left( T_{\bar 1,1}T_{[m]}^{(X)}T_{[n]}^{(Y)}\right)\right)^2
%\chi_{R_1,R_2}^R\left( T_{\bar 1,1}T_{[m]}^{(X)}T_{[n]}^{(Y)}\right)
\chi_{R_1,R_2}^R\left(\Omega\right)
\end{align}
We now expand $T_{\bar 1,1}=T_{2}^{(X,Y)}-T_{2}^{(X)}-T_{2}^{(Y)}$ so that
\begin{align}
T_{\bar 1,1}T_{[m]}^{(X)}T_{[n]}^{(Y)}=
T_{2}^{(X,Y)}T_{[m]}^{(X)}T_{[n]}^{(Y)}-T_{2}^{(X)}T_{[m]}^{(X)}T_{[n]}^{(Y)}-T_{[m]}^{(X)}T_{2}^{(Y)}T_{[n]}^{(Y)}
\end{align}
We also have (see \emph{e.g.} \cite{quivcalc})
\begin{align}\label{chi omega}
\chi_{R_1,R_2}^R\left(\Omega\right)&=
\chi_{R_1,R_2}^R\left(\sum_{\sigma\in S_{m+n}}\sigma N^{C_\sigma}\right)=
\frac{g\,d_{R_1}d_{R_2}}{d_R}(n+m)!\,\text{Dim}_N(R)
\end{align}
Eq. \eqref{chi initial form} simplifies then to
\begin{align}\label{char exp}
&\la \cO\cO^\dagger\ra=
\frac{1}{m!n!}\,
\sum_{R_1\vdash m\atop R_2\vdash n}\,\sum_{R\vdash m+n}\,
\frac{1}{d_{R_1}\,d_{R_2}\,g}\,\text{Dim}_N(R)\,\left(
\chi_{R_1,R_2}^R\left( T_{\bar 1,1}T_{[m]}^{(X)}T_{[n]}^{(Y)}\right)\right)^2 
\end{align}
On the other hand, as shown in Appendix \ref{Appendix: two point generating function}
\begin{equation}\label{chiR full form}
\begin{array}{ll}
&\chi_{R_1,R_2}^R\left( T_{\bar 1,1}T_{[m]}^{(X)}T_{[n]}^{(Y)}\right)=\nn
&=\left\{
\begin{array}{l}

%g\,\,\chi_{R_1}(T_{[m]}^{(X)})\,\chi_{R_2}(T_{[n]}^{(Y)})\left[
(-1)^{c_{R_1}+c_{R_2}}\,g\,(m-1)!(n-1)!\left[
\frac{ \chi_R(T_{2}^{(X,Y)}) }{d_R}-
\frac{ \chi_{R_1}(T_{2}^{(X)}) }{d_{R_1}}-
\frac{ \chi_{R_2}(T_{2}^{(Y)}) }{d_{R_2}}
\right];\quad R_1,\,R_2 \text{ hooks}\nn
0\quad \text{otherwise}
\end{array}
\right.
\end{array}
\end{equation}
Here $c_{R_i}$ is the number of boxes in the first column of the Young diagram associated with the representation $R_i$. This expression restricts the sums over representations $R_1\vdash m,\,R_2\vdash n$ in \eqref{char exp} to a sum over hook representations $h_1\vdash m,\,h_2\vdash n$.

We now need an equation for $g(h_1,h_2;R)$, with $h_1$ and $h_2$ hook representations of $S_m$ and $S_n$ respectively. We specify any representation $R$ by the sequence of pairs of integers $R =((a_1,b_1),(a_2,b_2),\allowbreak...(a_d,b_d))$. In a Young diagram interpretation, $a_j$ ($1\leq j\leq d$) is the number of boxes to the right of the $j$-th diagonal box, and $b_j$ is the number of boxes below the $j$-th diagonal box. We refer to $d$ as the `depth' of the representation $R$. Let us write $h_1=(k_1,l_1)$, $h_2=(k_2,l_2)$ and $R=((a_1,b_1),(a_2,b_2))$. In Appendix \ref{Appendix: LR rule for hooks} we show that 
\begin{align}\label{Eq: LR for hooks short}
&g(h_1,h_2;R)\,=\,
%hook
\delta_{k_1+k_2,a_1}\,\delta_{l_1+l_2+1,b_1}\,\delta_{-1,a_2}\,\delta_{0,b_2}
%hook
+\delta_{k_1+k_2+1,a_1}\,\delta_{l_1+l_2,b_1}\,\delta_{0,a_2}\,\delta_{-1,b_2}\nn
&+
\sum_{\epsilon_1,\epsilon_2=0}^1\,\sum_{i=\epsilon_1\bar\epsilon_2}
^{\min(k_1-\bar\epsilon_1\bar\epsilon_2,k_2-\epsilon_1\epsilon_2)}\,\,\,\,\sum_{j=\bar\epsilon_1\epsilon_2}
^{\min(l_1-\bar\epsilon_1\bar\epsilon_2,l_2-\epsilon_1\epsilon_2)}
\delta_{k_1+k_2-i+\bar\epsilon_1\epsilon_2 ,a_1}\,
\delta_{l_1+l_2-j+\epsilon_1\bar\epsilon_2, b_1}\,
\delta_{i-\epsilon_1\bar\epsilon_2, a_2}\,
\delta_{j-\bar\epsilon_1\epsilon_2, b_2}
\end{align}
where $\bar \epsilon_{1,2}=1-\epsilon_{1,2}$. Using this identity, in Appendix \ref{Appendix: two point generating function} we derive the formula
\begin{align}\label{2pt func main text}
\langle&\Tr(X^mY^n)\Tr(X^mY^n)^\dagger\rangle\nn
&=\sum_{k_1,l_1=0}^m\,\sum_{k_2,l_2=0}^n\,\,\sum_{a_1,b_1=0\atop a_2,b_2=0}^{n+m}\,g\,\,\delta(k_1+l_1-m)\,\delta(k_2+l_2-n)\,\,\,F(a_1,b_1,a_2,b_2,k_1,l_1,k_2,l_2) 
\end{align}
where we defined the function
\begin{align}\label{F function definition main text}
&F(a_1,b_1,a_2,b_2,k_1,l_1,k_2,l_2)= 
\frac{k_1! k_2! l_1! l_2!\,(a_1-a_2) (b_1-b_2) }{4(a_1+b_2+1) (a_2+b_1+1)(k_1+l_1+1) (k_2+l_2+1)}\nn
&\times\binom{a_1+b_1}{b_1} \binom{a_2+b_2}{b_2} \binom{N+a_1}{a_1+b_1+1} \binom{N+a_2}{a_2+b_2+1}\times\nn
&\times
((a_1+b_1+1)(a_1-b_1)+(a_2+b_2+1)(a_2-b_2)+\nn
&
\qquad\qquad\qquad
-(k_1+l_1+1)(k_1-l_1)- (k_2+l_2+1)(k_2-l_2))^2
\end{align}
In \cite{Bhattacharyya:2008xy} a closed form for the two point function has been given by using a different approach based on Young-Yamanouchi symbols. We have checked agreement of \eqref{2pt func main text} with that closed form for up to $n=m=10$.
It is an interesting exercise to simplify \eqref{2pt func main text} into the closed form obtained in \cite{Bhattacharyya:2008xy}. 
It will also be interesting to apply the present franework to obtain formulae analogous to \eqref{2pt func main text} for more general GIOs corresponding to central elements of $ \cA ( m , n ) $.

In this section we have shown how to calculate a particular two point function of a central operator, without explicitly constructing projectors. The result rather follows from knowing how central operators of interest are generated via the star product of pure $X$ gauge invariants, pure $Y$ gauge invariants and descendants of half-BPS operators.

\subsection{Coloured Ribbon graphs}
The correlator computations above can be expressed in terms of ribbon graphs, equivalently the usual double-line graphs of large $N$ expansions, but with edges coming in two colors, as explained for example in \cite{KR10}. The graphs can be organised by the minimum genus of 
the surface they can be embedded in and these graphs of a given genus contribute to a fixed power of $N$. For small $ m ,n $, we have checked with GAP that directly computing the permutation sums for a given genus agree with the analytic result \eqref{2pt func main text} we have derived.

\section{Conclusions and future directions }\label{conclusions} 

In this paper, we initiated a systematic study of permutation centralizer algebras, in connection with gauge invariant operators. We focused our attention on the algebras $ \cA ( m ,n ) $ which are related to restricted Schur operators studied in the context of giant gravitons in AdS/CFT. Other closely related algebras are related to the Brauer basis for multi-matrix invariants, the covariant basis and to tensor models.

While many of the key formulae we have used were already understood in the literature on giant gravitons, we have emphasized the intrinsic structure of $ \cA ( m ,n ) $ as an associative algebra with a non-degenerate pairing. This means that it has a Wedderburn-Artin decomposition, which gives a basis for the algebra in terms of matrix-like linear combinations. The construction of these matrix-units in terms of representation theory data from $ S_{m+n } , S_m , S_n $ has already been extensively used in the context of giant gravitons, although the link to the Wedderburn-Artin decomposition has not been made explicit before. In addition to explicating this link, the new emphasis in this paper has been on the structure of the centre $ \cZ ( m ,n ) $ and the maximally commuting sub-algebra $ \cM ( m ,n )$. 
 
We have used the structure of $ \cM ( m ,n )$ as a polynomial algebra over $ \cZ ( m ,n ) $ to characterize the minimal number of charges needed to identify any 2-matrix gauge invariant (section \ref{charges-result}). It will be interesting to  generalize this discussion to gauge invariants for more general gauge groups. 
 
Two key structural facts about $ \cA ( m ,n ) $ have played a role in the computation of correlators in Section \ref{sec:finiteNcorrelator}. The first is that $ ( x^m) * (y^n) = ( x^m y^n)$ and the second is that $ (x^my^n)$ is part of $ \cZ ( m ,n ) $. The non-degenerate pairing on $ \cA ( m ,n ) $, when restricted to elements in the centre, can be expressed in terms of characters of $ S_n , S_m , S_{ n+m}$ without requiring more detailed representation theory data such as matrix elements and branching coefficients. These are in general computationally hard to calculate, although there has been progress in the context of ``perturbations of half-BPS giants''. This makes it very interesting to understand the structure of the centre $ \cA ( m , n )$. A special case is $ \Z [ \mathbb{C} [ S_n ] ] $, which is the algebra of class sums in $ S_n$.

\subsection{Structure of the centre }

A number of questions about $ \cA ( n ) $, $ \cA ( m , n )$ and the centre $ \cA ( m ,n ) $ can be explored experimentally, with the help of group theory software, notably GAP. In particular, since $ \cZ ( m , n ) $ is generated by the centre of $S_m $, the centre of 
 $S_n$ and that of $ S_{ n + m } $ it is a useful first step to know about these centres. 
 
Since $ S_n$ is generated by transpositions, one might naively expect that the sum of permutations $T_2$ will generate $ \cA ( n )$. This is actually not true. We know that $T_2$ obeys a relation of degree $p(n)$
\begin{align} 
\prod_{ R \vdash n } \left ( T_2 - { \chi_R ( T_2) \over d_R } \right ) =0 
\end{align}
If this is the only relation, then we know that $T_2$ alone generates $ \Z [ \mathbb{C} [ S_n ] ] $. However simpler relations occur when there are coincidences in the normalized characters, e.g. two different irreps have the same normalized character. In fact the the failure of $T_2$ to generate centre is always correctly predicted by the degeneracies of the normalized characters. If we take 
\begin{align} 
\prod _R {}^{{}^{{}^\prime}} ( T_2 - { \chi_R ( T_2) \over d_R } ) =0 
\end{align}
where the product is taken over a maximal set of irreps with distinct normalized characters, we are getting an element in $ \mC [ S_n]$ which vanishes in all irreps. It is a central element, so the matrix elements in any irrep are proportional to the identity. We conclude that the above element vanishes. Given that the Peter-Weyl theorem gives an isomorphism between $ \mC [S_n ] $ and matrix elements of irreps, it follows that something which has vanishing matrix elements in all irreps should be identically zero.

% In fact this works for any class sum, $T_p$. 
% We have done computer checks showing that the Span of 
% central elements generated by powers of $T_p$ is correctly predicted by the above argument. 
% These powers are computed using the $ C_{ i j k } $ for the centre of $S_n$, 
% which are computed using the sum of a product of 3 characters. 
% ( THIS IS NICE BUT MAYBE NOT NEEDED). 

Even for large $n$, it is possible to check that the centre of $\mathbb C[S_n]$ is generated by a small number of $T_{p}$'s. Using GAP we tested that $T_{[2,1^{n-2}]}$ and $T_{[3,1^{n-3}]}$ are enough to generate the centre for $\mathbb C[S_n]$ up to $n=14$. The procedure we used to perform these checks is the following.
We know that the set of projectors $\{P_R\}$, with $R$ integer partition of $n$, generate the centre of $S_n$. We can compute the overlap of $P_R$ with the $k$-th power of $T_{p}$, that we simply write as $T_{p}^k$: 
\begin{align}
\lara{T_{p}^k,P_R} &= \delta(T_{p}^a\,P_R) = 
\frac{1}{n!}\sum_{S\vdash n} \chi_S(T_{p}^k)\chi_S(P_R)
=\frac{1}{n!}\sum_{S\vdash n}d_S\left(\frac{\chi_S(T_{p})}{d_S}\right)^k\,\chi_S(P_R)\nn
&=d_R\left(\frac{\chi_R(T_{p})}{d_R}\right)^k
\end{align}
Similarly, we can derive
\begin{align}\label{overlaps T2T3 centre}
\lara{T_{p}^kT_{q}^l,P_R} &= d_R\left(\frac{\chi_R(T_{p})}{d_R}\right)^k\left(\frac{\chi_R(T_{q})}{d_R}\right)^l
\end{align}
Now we construct the $AB \times p(n)$ matrix $M(A,B)$, whose matrix elements are the overlaps \eqref{overlaps T2T3 centre}:
\begin{align}
\left. M(A,B)\right|_{(k,l) , R} = d_R\left(\frac{\chi_R(T_{p})}{d_R}\right)^k\left(\frac{\chi_R(T_{q})}{d_R}\right)^l
\end{align}
with $0\leq k <A$ and $0\leq l <B$. By computing the rank of this matrix we obtain the number of independent central elements in $\mathbb C[S_n]$ that are obtained by taking at most $A-1$ powers of $T_{p}$ and $B-1$ powers of $T_{q}$. This method can be easily generalised to obtain the number of central elements generated by the string of operators $T_{p_1}^{k_1}T_{p_2}^{k_2}\cdots T_{p_N}^{k_N}$.

These studies on the centre of $\mathbb C[S_n]$ inspire a similar analysis for centre of $\A(m,n)$. The task is to find a minimal set of generators for $\Z(m,n)$ as a polynomial algebra. The importance of this problem is discussed in section \ref{charges-result}. Concretely, we would like to determine $ N^{min} ( \Z ( m ,n ) ) $. 
 There are many approaches one can take in this case, which would be interesting to investigate in the future. For example, using GAP we checked that low powers of the sum of two- and three-cycles permutations, $T_2^{(m+n)}$ and $T_3^{(m+n)}$, together with the generators of the centres of $\mathbb C[S_m]$ and $\mathbb C[S_n]$, generate the whole centre $\Z(m,n)$. We leave a more systematic discussion of this problem for future work.

\subsection{ Construction of quarter-BPS operators beyond zero coupling and the structure constants of $ \cA(m,n)$. }

The centre of $ \mC [ S_{ m+n} ] $ is denoted by $ \cZ [ \mC [ S_{ m+n} ]] $. 
$ \cZ [ \mC [ S_{ m+n} ] ] $ is a commutative sub-algebra of $ \cA ( m , n ) $. 
 The $ \cA (m , n)$ algebra is a module over $ \cZ[\mC [ S_{ m+n} ]] $. We can write
\begin{align} 
\cT_p N_i = \tilde C_{ p j }^{ k } N_{ k} 
\end{align}
for some coefficients $\tilde C$. The $ \cT_p$ are themselves linear combinations of 
necklaces: 
\begin{align} 
\cT_p = T_p^i N_i 
\end{align}
Hence 
\begin{align} 
\cT_p N_i = T_p^j N_j N_i = T_p^j C_{jk}^l N_l 
\end{align} 

Another subspace in $ \cA ( m , n ) $ is the subspace of symmetrised traces. A symmetrised trace $S_v$ can be parametrised by a {\it vector partition} $v $ of $ ( m , n )$. We can expand $S_v$ on the basis of necklaces $\{N_k\}$ as
\begin{align} 
S_v = S_v^k N_k 
\end{align}
Symmetrised traces and their products are quarter-BPS at weak coupling in the large $N$ limit. 
One can get the complete set of $ 1/N$ corrected BPS states at large $N$ by acting on $S_v$ with $ \Omega^{-1}$ which belongs to $ \cZ [ \mC [ S_{ m+n} ] ] \otimes \mC( 1/N ) $
\cite{BHR1,BHR2,Brown:2010pb,quivcalc}. The coefficients of $T_p $ are easily computable. The expansion of $T_p$ in terms of necklaces is also easily computable. The non-trivial part of the calculation is the $C_{ij}^k$ of the necklace algebra $ \cA(m ,n)$. For any symmetrised trace $S_v$, the corrected operator is 
\begin{align} 
\Omega^{-1}_k S_v = \Omega^{-1 }_p T_p S_v^j N_j = \Omega^{-1}_p S_v^j \tilde C_{ p j }^k N_k =
 \Omega^{-1}_p S_v^j T_p^l C_{l j }^k N_k 
\end{align}

\subsubsection{ Central quarter BPS sector } 

A subspace of symmetrised trace elements is central. The symmetrised trace elements 
give a subspace of $ \cA ( m ,n ) $ and the central elements form another subspace. The intersection is the space of central symmetrised traces. The dimension of this subspace can be computed for small $ m ,n $ using GAP. Suppose $S^C $ is an element in this subspace. Then elements $\Omega^{-1} S^C$ in $ \cA ( m ,n ) $ are very interesting. They are quarter-BPS beyond zero coupling and they are central, so computations of their correlators have the simplicity of the centre. The computations can be done using knowledge of the characters of $S_m , S_n , S_{ m+n} $, without knowing branching coefficients. From AdS/CFT this central quarter BPS sector should have a dual in the space-time theory, \emph{e.g.} some sub-class of states in the tensor product of super-graviton states. An interesting question is to compute their correlators in space-time and verify the matching with the gauge theory computations.

\subsection{ Non-commutative geometry and topological field theory }

Studies in non-commutative geometry in string theory suggest that open strings can be associated to non-commutative algebras and the centre is related to closed strings \cite{MooreSegal}. If we apply this thinking to $ \cA ( m , n ) $ and $ \cZ ( m , n )$,
how do we interpret these emergent open and closed strings? The traditional view is that Yang-Mills theory is the open-string picture in AdS/CFT with the closed string picture given by the AdS description, so this is an intriguing question. Non-commutative algebras and their centre have also been discussed in non-commutative geometry in \cite{Berenstein:2001jr}. The study of the pair $ \{ \cA ( m ,n ) , \cZ ( m ,n ) \} $ should form an interesting example of this discussion. Additionally we have the Cartan $ \cM (m , n )$ here, with physical relevance in distinguishing the multiplicity labels. So a more complete picture of strings and non-commutative geometry 
for the triple $ \{ \cA ( m ,n ) , \cM ( m ,n ) , \cZ ( m ,n ) \}$ looks desirable. Given that the infinite direct sum $ \cA ( \infty , \infty )$ comes up 
in connection with matrix invariants, it would also be interesting to study the triple $ \{ \cA ( \infty ,\infty ) , \cM ( \infty ,\infty ) , \cZ ( \infty ,\infty ) \}$
from this point of view. Some relevant  work in this direction is in \cite{Kimura:2014mka}.

\subsection{ Other examples of permutation centralizer algebras and correlators } 

Based on our study of $ \cA ( m ,n ) $, we outline some properties of the other examples of permutation centralizer algebras given in section \ref{sec:def}
and sketch the connection to correlators. 
 We leave a more detailed development for the future. 

Consider $ \cB_N ( m ,n ) $, which is the subspace of the Brauer algebra $ B_N ( m ,n ) $ 
invariant under $ \mC [ S_m \times S_n ]$. This is Example 3 in Section \ref{sec:def}. Brauer algebras were used to construct gauge invariant operators in \cite{KR1} from tensor products of a complex matrix and its conjugate. For an element $ b $ in the walled Brauer algebra $ B_N ( m ,n ) $, we use 
\bea 
tr_{ m ,n } \left ( Z^{ \otimes m } \otimes \bar Z^{ \otimes n } b \right ) 
\eea 
where the trace is taken in $ V^{ \otimes m } \otimes \bar V^{ \otimes n }$, a tensor product of 
fundamentals and anti-fundamentals of $U(N)$. We focus here on the case $ m+n \le N $. 
The number of gauge invariant operators is 
\bea\label{numBraOp} 
\sum_{ \gamma , \alpha , \beta } ( M^{ \gamma}_{ \alpha , \beta })^2
\eea
where $ \gamma $ labels an irrep of $B_N ( m ,n ) $, while $ \alpha , \beta $ are irreps of $ S_m $ and $S_n$ respectively. $M^{ \gamma}_{ \alpha , \beta }$ is a multiplicity with which $ ( \alpha , \beta )$ appears in the reduction of $ \gamma $ from $ B_N ( m , n ) $ to its $ \mC [ S_m \times S_n ] $ sub-algebra. The sum of squared dimensions in \eqref{numBraOp} is the dimension of the algebra $ \cB_N ( m , n ) $. This is a non-commutative algebra. The dimension of its centre is the number of triples $ ( \gamma , \alpha , \beta ) $ for which $M^{ \gamma}_{ \alpha , \beta }$ is non-vanishing. There is a maximally commuting sub-algebra of dimension equal to the sum 
\bea
\sum_{ \gamma , \alpha , \beta } M^{ \gamma}_{ \alpha , \beta }
\eea
This follows since the $ ( \gamma , \alpha , \beta ) $ give a Wedderburn-Artin decomposition of $ \cB_N ( m , n ) $. A tractable sector of correlators should be given by the centre of $ \cB_N ( m , n ) $ and more detailed study of the structure of this centre will be useful.

The next algebra of interest is the sub-algebra $ \cK ( n )$ of $ \mC [S_n ] \times \mC [S_n ]$ which is invariant under conjugation by $ Diag ( \mC [ S_n ] ) $. Let us denote this 
as $ \cA_{ diag } ( n , n )$. We can generate elements in this algebra by summing over the elements of the sub-group
\bea 
\sigma_1 \otimes \sigma_2 \rightarrow \sum_{ \gamma \in S_n } \gamma \sigma_1 \gamma^{-1} \otimes 
\gamma \sigma_2 \gamma^{-1} 
\eea 
The dimension of this algebra is 
\bea 
\sum_{ R , S , T } C ( R , S , T )^2 
\eea
where $ C ( R , S , T )$ is the Kronecker coefficient, i.e. the number of times the irrep $T$ of $ S_n$ appears in the tensor product $ R \otimes S$. 
The dimension of the centre is the number of triples $ ( R , S , T) $ for which the $ C ( R , S , T )$ is non-zero. A maximal commuting sub-algebra has dimension 
\bea 
\sum_{ R , S , T } C ( R , S , T )
\eea
These properties follow from the fact the Wedderburn-Artin decomposition of the algebra $ \cK ( n ) $ has blocks labelled by triples $ ( R , S , T )$ with non-vanishing $ C ( R , S , T )$. 
An explicit formula for this decomposition is 
\bea 
Q^{ R , S , T }_{ \tau_1 , \tau_2 } =\sum_{ \sigma_1 , \sigma_2 } \sum_{ i_1 , i_2 , i_3 , j_1 , j_2 } 
S^{R , S , T , \tau_1 }_{ i_1 , i_2 , i_3 } S^{R , S , T , \tau_2 }_{ j_1 , j_2 , i_3 } 
 D^{ R }_{ i_1 j_1} ( \sigma_1 ) D^S_{ i_2 j_2 } ( \sigma_2 ) ~ \sigma_1 \otimes \sigma_2 
\eea
The $D$'s are representation matrices for $S_n$ irreps. The $S$'s are Clebsch-Gordan coefficients. One verifies, using equivariance properties of the Clebsch's that these are invariant under conjugation by the diagonal $S_n$. 

There is another definition of $ \cK( n ) $ which is more symmetric in $ ( R , S , T )$.
$ C ( R , S , T ) $ is also the multiplicity of invariants of the diagonal $S_n $ acting on $ R \otimes S \otimes T$. $ \cK( n )$ can be defined as the subalgebra of $ \mC [S_n ] \otimes \mC [S_n ] \otimes \mC [S_n ] $ which is invariant under left action by the diagonal $ \mC [ S_n]$ and right action by the diagonal $ \mC [S_n ]$. These invariant elements can again be constructed by averaging 
\bea 
\sum_{ \gamma_1 , \gamma_2 } ( \gamma_1 \sigma_1 \gamma_2 , \gamma_1 \sigma_2 \gamma_2 , \gamma_1 \sigma_3 \gamma_2 ) 
\eea
A representation basis is given by 
\bea 
\sigma_1 \otimes \sigma_2 \otimes \sigma_3 
D^{R}_{ i_1 , j_1 } ( \sigma_1 ) D^{S}_{ i_2, j_2 } ( \sigma_2 ) D^{T}_{ i_3 , j_3 } ( \sigma_3 )
 S^{R , S , T , \tau_1 }_{ i_1 , i_2 , i_3 } S^{R , S , T , \tau_2 }_{ j_1 , j_2 , j_3 } 
\eea
labelled by $ R , S , T , \tau_1 , \tau_2$. 

These triples of permutations $ ( \sigma_1 , \sigma_2 , \sigma_3 ) $, with equivalences given by left and right diagonal action have appeared in the enumeration invariants for tensor models built from 3-index tensors \cite{tenscor}. The simplification from a description in terms of permutation triples to one in terms of permutation pairs was also described there, which lead to a connection between 3-index tensor invariants and Belyi maps. By analogy with the discussion in this paper, we expect that the centre of $ \cK ( n )$ will lead to a class of simpler correlators in tensor models. The discussion of $ \cA ( \infty , \infty ) $ will analogously lead to
\bea 
\cK ( \infty ) = \bigoplus_{ n =0}^{ \infty } \cK ( n )
\eea
This space will have two products: one related to the algebra structure of $ \cK ( n )$ and 
one related to the multiplication of tensor invariants. Somewhat related algebraic structures appear in \cite{MMN13} and it would be useful to better understand these relations. As a last remark, consider the Kronecker multiplicities $ C ( R , R , T ) $, \emph{i.e.} in the special case where $ R = S $. These have also appeared in the construction of gauge-invariant multi-matrix operators in a basis which is covariant under the global symmetries \cite{BHR1,BHR2}. The structure of $ \cK ( n ) $ can thus also be expected to have implications for multi-matrix correlators in the covariant basis.

\vskip2cm

\begin{centerline}
{ \bf Acknowledgements} 
\end{centerline} 

\vskip.4cm 

We thank David Berenstein, Robert de Mello Koch and Edward Hughes for useful discussions, and Robert de Mello Koch for comments on an earlier draft.
 SR is supported by STFC consolidated grant ST/L000415/1 ``String Theory, Gauge Theory \& Duality." SR thanks the Simons Summer workshop 2015 for hospitality while part of this work was done. PM is supported by a Queen Mary University of London studentship.

\vskip2cm

\begin{appendix}

\section{Analytic formula for the dimension of $ \cM ( m , n)$ }\label{sec:AnalDim}

In this section we derive a formula for the dimension of $ \cM ( m , n ) $. 
This dimension is equal to the sum of Littlewood-Richardson coefficients
\bea 
\hbox{ Dim } \cM ( m , n ) = \sum_{ R_1 \vdash m , R_2 \vdash n } 
\sum_{ R \vdash m +n } 
g ( R_1 , R_2 , R ) 
\eea
The sum of squares of the Littlewood-Richardson coefficients is the dimension of $ \cA ( m ,n ) $ and has a simple 2-variable generating function. It is natural to ask if we can write 
a nice generating function for the dimension of $ \cM ( m ,n ) $. While we have not been able to derive something of comparable simplicity, we will derive two interesting expressions \eqref{pairing-formula} and \eqref{Dim(M) app} in terms of multi-variable polynomials. 

Let $T_p$ denote a conjugacy class of permutations with cycle structure determined by a vector $ (p_1, p_2 , \cdots )$, i.e. permutations with $p_i$ cycles of length $i$. Let now $\sigma_p$ be an element in $T_p$. For $ \sigma_p \in T_p$, it is known that \cite{macdonald1968theory}
%We will also write $ \chi_R ( T_p ) $ for $ \chi_R ( \sigma ) $ for any $ \sigma \in T_p$. \td{use something like $\sigma_p$, fix throughout}
\bea 
\sum_{R} \chi_R ( \sigma_p ) = \prod_i \hbox{ Coeff } \left ( f_i ( t_i ) , { t_i^{ p_i} \over p_i! } \right ) 
\eea
where 
\bea 
f_i ( t_i ) = e^{ { ( 1 - (-1)^i ) \over 2 } t_i + {i t_i^2 \over 2 } } 
\eea

We can define 
\bea 
F ( t_1 , t_2 , \cdots ) = \prod_i f_i (t_i) 
\eea
and write 
\bea 
\sum_{ R} \chi_R ( \sigma_p ) = \hbox{ Coeff } \left ( F ( t_1 , t_2 , \cdots ) , \prod_i { t_i^{ p_i} \over p_i! } \right ) 
\eea
It is also useful to define 
\bea\label{alsouseful} 
\tilde f_i ( t_i ) & = & f_i \left({ t_i \over i } \right) \cr 
\widetilde F ( t_1 ,t_2 , \cdots ) & = & F \left( t_1 , { t_2 \over 2 } , { t_3 \over 3 } \cdots \right) = F \left( \{ { t_i \over i } \} \right) \cr 
 & = & \prod_{ i : \rm{ odd} } e^{ t_i \over i } \prod_{ i =1 }^{ \infty} e^{ i t_i^2 \over 2 i } 
\eea
We can write the LR coefficients in terms of $T_p$'s as
\bea 
&& g ( R_1 , R_2 , R ) = { 1 \over m! n! } 
\sum_{ \sigma_1 \in S_{ m } } \sum_{ \sigma_2 \in S_{ n} }
\chi_{ R_1} ( \sigma_1 ) \chi_{ R_2} ( \sigma_2 ) \chi_R ( \sigma_1 \circ \sigma_2 ) \cr 
&& = \sum_{ p \vdash m } \sum_{ q \vdash n } 
 \chi_{ R_1} ( T_{ p } ) \chi_{ R_2} ( T_{ q }) \chi_{ R} ( T_{ p } \circ T_{q } ) 
\prod_{ i } { 1 \over i^{ p_i + q_i } p_i! q_i! } \cr 
&& 
\eea
This uses the fact that the number of permutations in the class $T_p$ is $ n!/ \prod_i i^{ p_i} p_i!$. Now use the above formula for $\sum_R \chi_R ( T_p) $, to obtain 
\bea 
&& \sum_{ R _1 , R_2 , R } g ( R_1 , R_2 , R ) \cr 
&& = \sum_{ p \vdash m } \sum_{ q \vdash n } \prod_i \hbox{ Coeff} ( \tilde f_i ( s_i ) , { t_i^{ p_i} } ) \hbox{ Coeff } ( \tilde f_i ( t_i ) , { t_i^{ q_i} } )
\hbox{ Coeff } ( f_i ( u_i ) , { u_i^{ p_i + q_i} } ) ( p_i + q_i )! \cr 
&& = \sum_{ p \vdash m } \sum_{ q \vdash n } 
\hbox{ Coeff } ( \widetilde F ( \vec s ) \widetilde F ( \vec t ) F ( \vec u ) , \prod_i s_i^{ p_i} t_i^{ q_i} u_i^{ p_i + q_i} ( p_i+q_i)!^{-1} ) \cr 
&& = \sum_{ p \vdash m } \sum_{ q \vdash n } 
\hbox{ Coeff } ( \widetilde F ( \vec s ) \widetilde F ( \vec t ) \widetilde F ( \vec u ) , \prod_i s_i^{ p_i} t_i^{ q_i} u_i^{ p_i + q_i} ) i^{ p_i + q_i} ( p_i + q_i ) ! 
\eea
It is useful to make the substitutions $ s_i \rightarrow s^i z_i , t_i \rightarrow t^i z_i , u_i \rightarrow \bar z_i $ and to introduce a pairing \footnote{ Alternatively we can think about expectation values in a Fock space with $ z_i \rightarrow a_i , \bar z_i \rightarrow a^{ \dagger}_i$. This would allow us to write the subsequent formulae in terms of quantities in a 2D field theory. This perspective could be fruitful, but we will leave its exploration for the future} 
\bea 
\langle z_j^{ k} , \bar z_i^l \rangle = \delta_{ ij} ~ \delta_{ kl } ~k!~ i^k 
\eea
With these substitutions define 
\bea\label{subsdef} 
\cF ( z_i , s ) = \widetilde F ( t_i \rightarrow s^i z_i ) 
\eea
Then we can write
\begin{equation}\label{pairing-formula} 
\boxed{ ~~~~ \hbox{ Dim } ( \cM ( m ,n ) ) = \langle \hbox{ Coeff } ( \cF ( z_i , s ) \cF ( z_i , t ) 
\cF ( z_i , u =1 ) , s^m t^n ) \rangle ~~~~ 
} 
\end{equation}
This has been checked for very simple cases, e.g. up to $ ( m , n ) = (3,3)$

\subsection{Multi-variable polynomials }

It is useful to isolate the multi-variable polynomials in the $z_i $ variables at each order in the $ s, t $ variables. 
Let us introduce the quantities
\bea 
&& \cA ( \vec z , s ) = \prod_i \exp \left [ { s^{ 2i } z_i^2 \over 2 i } \right ] \cr 
&& \cB ( \vec z , s ) = \prod_{ i = 1, 3, .. } \exp \left [ \frac{s^i z_i}{i} \right ] 
\eea
It follows from previous formulae \eqref{alsouseful} and \eqref{subsdef} that 
\bea 
 \cF ( \vec z , s ) = \cA ( \vec z , s ) \cB ( \vec z , s ) 
\eea
Introducing polynomials $\cF_m ( \vec z ) $ for each order in $s$ we can rewrite the latter quantity as
\bea \label{AB prod app}
 \cF ( \vec z , s ) = \sum_{ m =0 } \cF_m ( \vec z ) s^m 
\eea
We will now write formulae for the coefficients of $ s^m $ in $ \cA $ and $ \cB $. For $\cA( \vec z , s )$ we derive 
\begin{align}
 &\cA ( \vec z , s ) = \sum_{ m = 0 }^{ \infty } \cA_{ 2m } ( \vec z ) s^{ 2 m} = \sum_{ p_1 , p_2 , \cdots = 0 }^{ \infty } \prod_{ i=1}^{ \infty } { s^{ 2 i p_i } z_i^{ 2 i p_i } \over ( 2i )^{p_i} p_i ! }
\end{align}
so that 
\begin{align}
\cA_{ 2m } ( \vec z ) = \sum_{ p \vdash m } { z_i^{ 2 i p_i } \over (2i)^{ p_i } p_i! } 
\end{align}
We can also define $ \cA_m ( \vec z ) $ to be zero for odd $m$ and equal to the above for the even values. It is useful to define the coefficients of $ z_1^{2 p_1} z_2^{ 4 p_2} \dots z_i^{2ip_i} $ in the $ \cA ( \vec z , s=1 ) $ as 
\bea 
\cA_{ [ p ] } = \cA_{ [ p_1 , p_2 \cdots ] } = \prod_i { 1 \over p_i! ( 2i)^{ p_i} } 
\eea
so that we may write 
\bea 
\cA_{ 2 m } = \sum_{ p \vdash m } \cA_{ [ p ] } \prod_{ i =1}^{ \infty } z_i^{ 2 i p_i } 
\eea
Similarly, for $\cB ( \vec z , s )$ we obtain
\bea 
 \cB ( \vec z , s ) = \prod_{ i =0 }^{ \infty } \exp \left [ {s^{ ( 2i + 1 ) } z_{ 2i+1} \over ( 2i+1) } \right ] 
\eea
and
\bea 
\cB_{ m } ( \vec z ) = \sum_{ \{ p_1 , p_3 \cdots \} \vdash m } \prod_{ i ~ odd } { z_{ i }^{ i p_{ i } } \over ( i )^{ p_i } p_{ i} ! } 
\eea
Therefore it is natural to define 
\bea 
&& \cB_{ [ p_1 , p_3 , \cdots ] } = \prod_i { 1 \over i^{ p_i} p_i ! } \cr 
&& \cB_{ m } ( \vec z ) = \sum_{ p \vdash m } \cB_{ [ p_1 , p_3 , \cdots ] } \prod_{ i ~ odd } z_i^{i p_i} 
\eea
Going back to \eqref{AB prod app} we get, using the formulae just derived
\begin{align}
 \cF_{ m } ( \vec z ) &= \sum_{ k = 0 }^{ m } \cA_k ( \vec z ) \cB_{ m - k } ( \vec z )
=\sum_{k=0}^{ \lfloor { m \over 2 } \rfloor } \cA_{2k} ( \vec z ) \cB_{ m - 2k } ( \vec z )\nonumber\\[3mm]
&=\sum_{k=0}^{ \lfloor { m \over 2 } \rfloor } \sum_{r\vdash k}\sum_{q\vdash m-2k\atop q \text{ odd}} \cA_{[r]} \cB_{[q]} \prod_i z_i^{i(2r_i+q_i)}
\end{align}
Grouping terms with the same power of $z_i$ we obtain
\bea 
&& \cF ( \vec z , s = 1 ) = \sum_{ [ p_1 , p_2 ... ] } \cF_{ [ p_1 , p_2 ... ] } \prod_{ i } z_i^{ i p_i }
\eea
with
\bea
&& \cF_{ [p] } = \sum_{ [ r_1 , r_2, ... ] } \sum_{ [ q_1 , q_2 \cdots ] } \cA_{ [ r_1 , r_2 \cdots ] } \cB_{ [ q_1 , q_2 , \cdots ] } \prod_{ i ~ even } \delta ( p_i , 2 r_i ) 
~ \prod_{ i ~ odd } \delta ( p_i , 2 r_i + q_i ) 
\eea

Note that the function $ \cF ( \vec z , s ) $ is closely related to the generating function for the cycle 
indices of $ S_n $ which is 
\bea 
&& \cZ ( \vec z , t ) = \exp \left [ \sum_{ i =1}^{ \infty }{ t^i z_i \over i } \right ] \cr 
&& \tilde \cA ( \vec z , s ) = \left ( \cZ ( z_i \rightarrow z_i^2 , s \rightarrow s^2 ) \right )^{ 1/2} \cr 
&& \tilde \cB ( \vec z , s ) = \left ( \cZ (z_{ 2i+1} \rightarrow z_{ 2i+1} , z_{ 2i } \rightarrow 0 ) \right )^{ 1/2 } 
\eea
We can work with the same function if we change the pairing. With the pairing
\bea 
\lara{ z_i^{ k_i } , z_j^{ k_j } } ~ = ~ \delta_{ i , j } \delta_{ k_i , k_j } k_i! ~ i^{ k_i } 
\eea
we can write the above formulae as 
\bea 
&& { \rm { Dim }} ( \cM ( m ,n ) ) = \lara{ \cF_m ( \vec z ) \cF_n ( \vec z ), \cF_{ m + n } ( \vec z ) }
\eea 
or, equivalently,
\bea \label{Dim(M) app}
&& { \rm { Dim }} ( \cM ( m ,n ) ) = \sum_{ p \vdash m } \sum_{ q \vdash n } \cF_{ p_1 , p_2 \cdots }\cF_{ q_1 , q_2 , \cdots }\cF_{ p_1 + p_2 , q_1 + q_2 , \cdots } \prod_{ i } i^{ p_i + q_i } ( p_i + q_i ) ! \cr 
&& = \sum_{ p \vdash m } \sum_{ q \vdash n }\cF_p \cF_q \cF_{ p+q } Sym ( p+q ) 
\eea
This is eq. \eqref{dimension formula M}.

\section{LR rule for hook representations}\label{Appendix: LR rule for hooks}

Here we derive the LR decomposition rule for the tensor product of two hook representations.
Let us consider three representations $R,\,R_1$ and $R_2$ of $S_{m+n},\,S_m$ and $S_n$ respectively. The LR coefficient $g(R_1,R_2;R)$ gives the multiplicity with which the representation $R_1\otimes R_2$ appear in the representation $R$ upon its restriction to $S_m\times S_n$. There is a systematic procedure to obtain such coefficients \cite{FulHar}, that we now briefly review. %This is a diagrammatic procedure, that uses Young diagrams.
We take the Young diagrams corresponding to $R_1$ and $R_2$, and we start by decorating the latter as follows. We write `1' in all the boxes of the first row, `2' in all the boxes of the second row and so on in a similar fashion until the last row. Then we proceed to move all the `1' boxes from $R_2$ to $R_1$, ensuring that 
that we produce legal Young diagrams and no two copies of `1' appear  in the same column. We then move the `2' boxes following the same rules, and so on. 
In doing so, we also require a {\it reading condition}.   At any step, reading from right lo left along the first  row and then subsequent rows, 
 the number of `1' boxes must be greater or equal to  the number of `$2$' boxes. Similarly, the number of `2' boxes must be  greater or equal to the number of `$3$' boxes, and so on.

At the end of this procedure we are left with a collection of Young diagrams, made with $m+n$ boxes. If two or more of the resulting diagrams are identical (that is, the not only match in shape but also in the numbering of their boxes), we only retain one of them. Otherwise, if $k$ diagrams $R$ appear with the same shape but different numbering, we can say that $g(R_1,R_2;R)=k$. These will be the prescriptions that we will follow to derive our LR formula.

We specify any representation $R$ by the sequence of pairs of integers $R =((a_1,b_1),(a_2,b_2),\allowbreak...(a_d,b_d))$. In a Young diagram interpretation, $a_j$ ($1\leq j\leq d$) is the number of boxes to the right of the $j$-th diagonal box, and $b_j$ is the number of boxes below the $j$-th diagonal box. We refer to $d$ as the `depth' of the representation $R$. Hooks therefore are representations of depth 1. Schematically, in this appendix we will obtain the RHS of
\begin{align}
(k_1,l_1)\otimes (k_2,l_2)=\bigoplus\, ((a_1,b_1),(a_2,b_2))
\end{align}

In our derivation we imagine to keep the first hook fixed, and to add to it boxes coming from the second diagram. In doing so we are careful to follow the LR prescription. The boxes of the second diagram are decorated by a `1' or a `$v$', depending whether they come from the first row of the diagram or not. The tensor product $(k_1,l_1)\otimes (k_2,l_2)$ will decompose into a direct sum of a varying number of depth 2 representation and precisely two hooks (regardless of the actual value of $k_{1,2},\,l_{1,2}$). These hooks are
\begin{align}
&\text{Hook 1:}\quad (k_1+k_2+1,l_1+l_2)\nn
&\text{Hook 2:}\quad (k_1+k_2,l_1+l_2+1)
\end{align}
Notice that we can rewrite them using the notation we use for the depth two diagram as
\begin{align}
&\text{Hook 1:}\quad ((k_1+k_2+1,l_1+l_2),(0,-1))\label{App: hook1}\\[3mm]
&\text{Hook 2:}\quad ((k_1+k_2,l_1+l_2+1),(-1,0))\label{App: hook2}
\end{align}
This notation will be helpful at a later stage.

We now turn to the depth two representations. We proceed systematically, grouping them into four categories according to the two yes/no questions:
\begin{itemize}
\item[1)] Is there a $\young(1)$ in the first column of the resulting diagram?
\item[2)] Is there a $\young(v)$ in the first row of the inner hook of the resulting diagram?
\end{itemize}
We now analyse these four possibilities. 
%The hooks will fall in the class of diagrams that answer `No' to question 2 above: one will be included in the $(Y,N)$ set, the other in the $(N,N)$ set.

\subsection{(Y,Y) case}
The diagrams in this class are of the form
\begin{figure}[H]
\begin{center}\includegraphics[scale=3.7]{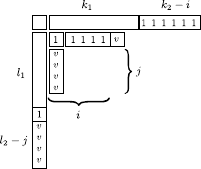}\\[1mm]
\caption{(Y,Y) case}\label{YYdiag}
\end{center}
\end{figure}
%They can be described by the expression
%\begin{align}
%(Y,Y):\qquad
%((k_1+k_2-i,l_1+l_2-j+1),(i,j-1))
%\end{align}
%The boundary for $i$ is
%\begin{align}
%0\leq i\leq \min(k_1,k_2-1)
%\end{align}
%The boundary for $j$ is
%\begin{align}
%1\leq j\leq \min(l_1+1,l_2)
%\end{align}
%We now write $j'=j-1$ so that (dropping the prime)
They can be described by the expression
\begin{align}
(Y,Y):\qquad
((k_1+k_2-i,l_1+l_2-j),(i,j))
\end{align}
where $i$ and $j$ are constrained by the boundaries
\begin{align}
&0\leq i\leq \min(k_1,k_2-1)\nn
&0\leq j\leq \min(l_1,l_2-1)
\end{align}
The upper bound on $i$ is $\min(k_1,k_2-1)$ because, if $k_1\geq k_2$, we cannot remove all the $k_2$ $\young(1)$ type boxes from the first row. This has to be avoided since by construction the rightmost box in the second row has to be a $\young(v)$ type box. A diagram with no $\young(1)$ type boxes on the first row and a $\young(v)$ type box at the end of the second row would violate the LR reading condition.
%The $j$ bound starts with a 1 because there has to be a $\young(v)$ in the second row.\\

\subsection{(Y,N) case}
The diagrams in this class are of the form
\begin{figure}[H]
\begin{center}\includegraphics[scale=3.7]{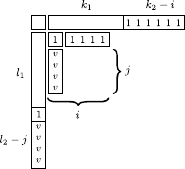}\\[1mm]
\caption{(Y,N) case}\label{YNdiag}
\end{center}
\end{figure}
They can be described by the expression
\begin{align}
(Y,N):\qquad
((k_1+k_2-i,l_1+l_2-j+1),(i-1,j))
\end{align}
%When $i=0$ the only meaningful diagram (a hook) is obtained for $j=0$.
with the boundaries
\begin{align}
&1\leq i\leq \min(k_1,k_2)\nn
&0\leq j\leq \min(l_1,l_2)
\end{align}
%\begin{equation}
%\left\{
%\begin{array}{l}
%1\leq i\leq \min(k_1,k_2)\nn
%
%0\leq j\leq \min(l_1,l_2)
%\end{array}
%\right\}
%\qquad\bigcup\qquad
%\{i=j=0\}
%\end{equation}
%The single point $i=j=0$ is the hook $((k_1+k_2,l_1+l_2+1),(-1,0))\equiv (k_1+k_2,l_1+l_2+1)$.

\subsection{(N,N) case}
The depth two diagrams in this class are of the form
\begin{figure}[H]
\begin{center}\includegraphics[scale=3.7]{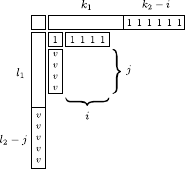}\\[1mm]
\caption{(N,N) case}\label{NNdiag}
\end{center}
\end{figure}
%\begin{align}
%(N,N):\qquad
%((k_1+k_2-i+1,l_1+l_2-j),(i-1,j))
%\end{align}
%When $i=0$ the only meaningful diagram (a hook) is obtained for $j=0$. Notice that we must also have $l_1>0$, otherwise we would have a $\young(v)$ in the second row. This is automatically enforced by the lower boundary on $j$
%The boundary for $i$ is
%\begin{align}
%0\leq i\leq \min(k_1,k_2+1)
%\end{align}
%The boundary for $j$ is
%\begin{align}
%0\leq j\leq \min(l_1-1,l_2)
%\end{align}
%As we stated before, this constraint automatically makes the set of diagrams with $l_1=0$ empty.
%We now write $i'=i-1$ so that (dropping the prime)
They can be described by the expression
\begin{align}\label{LR appendix: NN expression}
(N,N):\qquad
((k_1+k_2-i,l_1+l_2-j),(i,j))
\end{align}
with the boundaries
\begin{align}
&0\leq i\leq \min(k_1-1,k_2)\nn
&0\leq j\leq \min(l_1-1,l_2)
\end{align}
%However, the hook
%\begin{align}
%((k_1+k_2+1,l_1+l_2),(0,-1))\equiv ((k_1+k_2+1,l_1+l_2))
%\end{align}
%falls in this class too, and cannot be obtained smoothly from \eqref{LR appendix: NN expression}. We therefore have to manually add it to the $(N,N)$ set of diagrams.

\subsection{(N,Y) case}
The diagrams in this class are of the form
\begin{figure}[H]
\begin{center}\includegraphics[scale=3.7]{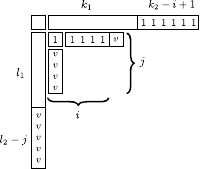}\\[1mm]
\caption{(N,Y) case}\label{NYdiag}
\end{center}
\end{figure}
These can be described by the equation
\begin{align}\label{NY equation}
(N,Y):\qquad
((k_1+k_2-i+1,l_1+l_2-j),(i,j-1))
\end{align}

The boundary for $i$ is
\begin{align}
0\leq i\leq \min(k_1,k_2)
\end{align}
The upper bound is $k_2$ and not $k_2+1$ because we cannot remove all the $\young(1)$ from the first row, as the rightmost box in the second row has to be a $\young(v)$ type box. In this way, we are enforcing the LR reading condition.
On the other hand, the boundary for $j$ is
\begin{align}\label{NY j range}
1\leq j\leq \min(l_1,l_2)
\end{align}
The lower bound is a 1 as by construction there has to be a $\young(v)$ box in the first row of the inner hook. 

% ---> the following is not necessary anymore!
%We can now write $j'=j-1$ so that (dropping the prime)
%\begin{align}
%(N,Y):\qquad
%((k_1+k_2-i+1,l_1+l_2-j-1),(i,j))
%\end{align}
%with the boundaries
%\begin{align}
%&0\leq i\leq \min(k_1,k_2)\nn
%&0\leq j\leq \min(l_1,l_2)-1
%\end{align}

\subsection{A summary}

These four cases comprise all possible valid depth two diagrams. Summarising our result, we have
\begin{itemize}
\item \textbf{ $(Y,Y)$ case:} \hspace*{2cm} $((k_1+k_2-i,l_1+l_2-j),(i,j))$
\begin{align}\label{YY summary}
&0\leq i\leq \min(k_1,k_2-1)\nn
&0\leq j\leq \min(l_1,l_2-1)
\end{align}
\vspace*{8pt}
\item \textbf{ $(Y,N)$ case:} \hspace*{2cm} $((k_1+k_2-i,l_1+l_2-j+1),(i-1,j))$
\begin{align}\label{YN summary}
&1\leq i\leq \min(k_1,k_2)\nn
&0\leq j\leq \min(l_1,l_2)
\end{align}
\vspace*{8pt}
\item \textbf{ $(N,N)$ case:} \hspace*{2cm} $((k_1+k_2-i,l_1+l_2-j),(i,j))$
\begin{align}\label{NN summary}
&0\leq i\leq \min(k_1-1,k_2)\nn
&0\leq j\leq \min(l_1-1,l_2)
\end{align}

\vspace*{8pt}
\item \textbf{ $(N,Y)$ case:} \hspace*{2cm} $((k_1+k_2-i+1,l_1+l_2-j),(i,j-1))$
\begin{align}\label{NY summary}
&0\leq i\leq \min(k_1,k_2)\nn
&1\leq j\leq \min(l_1,l_2)
\end{align}
\end{itemize}

We now introduce the boolean parameters
\begin{equation}
\epsilon_1=\left\{
\begin{array}{ll}\label{epsilon1}
0&\text{ If the answer to the first question is \textbf{no}}\\
1&\text{ If the answer to the first question is \textbf{yes}}
\end{array}
\right.
\end{equation}
and 
\begin{equation}
\epsilon_2=\left\{
\begin{array}{ll}\label{epsilon2}
0&\text{ If the answer to the second question is \textbf{no}}\\
1&\text{ If the answer to the second question is \textbf{yes}}
\end{array}
\right.
\end{equation}
With this notation we can compactly rewrite \eqref{YY summary} - \eqref{NY summary} as
\begin{align}
((k_1+k_2-i+\bar\epsilon_1\epsilon_2 ,l_1+l_2-j+\epsilon_1\bar\epsilon_2),(i-\epsilon_1\bar\epsilon_2,j-\bar\epsilon_1\epsilon_2))
\end{align}
where the sign $\bar{}$ denotes the logical negation of a boolean variable, so that $\bar\epsilon_{1,2}=1-\epsilon_{1,2}$. In this notation, $i$ and $j$ have the boundaries
\begin{align}
&\epsilon_1\bar\epsilon_2\leq i\leq \min(k_1-\bar\epsilon_1\bar\epsilon_2,k_2-\epsilon_1\epsilon_2)\nn
&\bar\epsilon_1\epsilon_2\leq j\leq \min(l_1-\bar\epsilon_1\bar\epsilon_2,l_2-\epsilon_1\epsilon_2)
\end{align}

By denoting $h_1=(k_1,l_1)$ and $h_2=(k_2,l_2)$, together with $R=((a_1,b_1),(a_2,b_2))$ we can then write
\begin{align}\label{Appendix eq: LR for hooks short}
&g(h_1,h_2;R)\,=\,
%hook
\delta_{k_1+k_2,a_1}\,\delta_{l_1+l_2+1,b_1}\,\delta_{-1,a_2}\,\delta_{0,b_2}
%hook
+\delta_{k_1+k_2+1,a_1}\,\delta_{l_1+l_2,b_1}\,\delta_{0,a_2}\,\delta_{-1,b_2}\nn
&+
\sum_{\epsilon_1,\epsilon_2=0}^1\,\sum_{i=\epsilon_1\bar\epsilon_2}
^{\min(k_1-\bar\epsilon_1\bar\epsilon_2,k_2-\epsilon_1\epsilon_2)}\,\,\,\,\sum_{j=\bar\epsilon_1\epsilon_2}
^{\min(l_1-\bar\epsilon_1\bar\epsilon_2,l_2-\epsilon_1\epsilon_2)}
\delta_{k_1+k_2-i+\bar\epsilon_1\epsilon_2 ,a_1}\,
\delta_{l_1+l_2-j+\epsilon_1\bar\epsilon_2, b_1}\,
\delta_{i-\epsilon_1\bar\epsilon_2, a_2}\,
\delta_{j-\bar\epsilon_1\epsilon_2, b_2}
\end{align}
where we also added the two hooks in the depth two notation, \eqref{App: hook1} and \eqref{App: hook2}. Explicitly, summing over the $\epsilon_{1,2}$ parameters, we get the lengthier expression
\begin{align}\label{Appendix eq: LR for hooks}
g&(h_1,h_2;R)=\nn
&=
%hook
\delta_{k_1+k_2,a_1}\,\delta_{l_1+l_2+1,b_1}\,\delta_{-1,a_2}\,\delta_{0,b_2}
+
\sum_{i=1}^{\min(k_1,k_2)}\,\sum_{j=0}^{\min(l_1,l_2)}
\delta_{k_1+k_2-i,a_1}\,
\delta_{l_1+l_2-j+1,b_1}\,
\delta_{i-1,a_2}\,
\delta_{j,b_2}
\nn
&+
%hook
\delta_{k_1+k_2+1,a_1}\,\delta_{l_1+l_2,b_1}\,\delta_{0,a_2}\,\delta_{-1,b_2}
+
\sum_{i=0}^{\min(k_1,k_2)}\,\sum_{j=1}^{\min(l_1,l_2)}
\delta_{k_1+k_2-i+1,a_1}\,
\delta_{l_1+l_2-j,b_1}\,
\delta_{i,a_2}\,
\delta_{j-1,b_2}
\nn
&+
\left(
\sum_{i=0}^{\min(k_1,k_2-1)}\,\sum_{j=0}^{\min(l_1,l_2-1)}+
\sum_{i=0}^{\min(k_1-1,k_2)}\,\sum_{j=0}^{\min(l_1-1,l_2)}
\right)
\delta_{k_1+k_2-i,a_1}\,
\delta_{l_1+l_2-j,b_1}\,
\delta_{i,a_2}\,
\delta_{j,b_2}
\end{align}
From this equation it is clear that $g(h_1,h_2;R)$ can be either 0, 1 or 2. In particular, $g(h_1,h_2;R)=2$ only if $R=((k_1+k_2-i,l_1+l_2-j),(i,j))$ and $0\leq i<\min(k_1,k_2) $, $0\leq j<\min(l_1,l_2) $.

\section{Deriving the two point correlator }\label{Appendix: two point generating function}
In this Appendix we will derive eq. \eqref{2pt func main text} from eq. \eqref{char exp}.
Let us start by considering the quantity
\begin{align}\label{chi app start}
\chi_{R_1,R_2}^R&\left( T_{\bar 1,1}T_{[m]}^{(X)}T_{[n]}^{(Y)}\right)
\end{align}
where we remind the reader that $R_1$, $R_2$ and $R$ are irreps of $S_m$, $S_n$ and $S_{m+n}$ respectively. Let us define $T_{2}^{(X,Y)}$, $T_{2}^{(X)}$ and $T_{2}^{(Y)}$ as the sum of transpositions in $S_{m+n}$, $S_m$ and $S_n$ respectively. We can expand \eqref{chi app start} as
\begin{align}\label{chi(R1,R2;R) exp}
\chi_{R_1,R_2}^R&\left( T_{\bar 1,1}T_{[m]}^{(X)}T_{[n]}^{(Y)}\right)\nn
&=
\chi_{R_1,R_2}^R\left(T_{2}^{(X,Y)}T_{[m]}^{(X)}T_{[n]}^{(Y)}\right)-
\chi_{R_1,R_2}^R\left(T_{2}^{(X)}T_{[m]}^{(X)}T_{[n]}^{(Y)}\right)-
\chi_{R_1,R_2}^R\left(T_{[m]}^{(X)}T_{2}^{(Y)}T_{[n]}^{(Y)}\right)\nn
&=%\displaybreak
g\frac{ \chi_R(T_{2}^{(X,Y)}) }{d_R}\,\chi_{R_1}(T_{[m]}^{(X)})\,\chi_{R_2}(T_{[n]}^{(Y)})-
\frac{1}{g\,d_{R_1}d_{R_2}}\chi_{R_1,R_2}^R(T_{2}^{(X)})\,\chi_{R_1,R_2}^R\left(T_{[m]}^{(X)}T_{[n]}^{(Y)}\right)+\nn
&
\qquad\qquad
-\frac{1}{g\,d_{R_1}d_{R_2}}\chi_{R_1,R_2}^R(T_{2}^{(Y)})\,\chi_{R_1,R_2}^R\left(T_{[m]}^{(X)}T_{[n]}^{(Y)}\right)\displaybreak\nn
&=
g\frac{ \chi_R(T_{2}^{(X,Y)}) }{d_R}\,\chi_{R_1}(T_{[m]}^{(X)})\,\chi_{R_2}(T_{[n]}^{(Y)})-
\frac{ \chi_{R_1}(T_{2}^{(X)}) }{d_{R_1}}\,\chi_{R_1,R_2}^R(T_{[m]}^{(X)}T_{[n]}^{(Y)})+\nn
&
\qquad\qquad
-\frac{ \chi_{R_2}(T_{2}^{(Y)}) }{d_{R_2}}\,\chi_{R_1,R_2}^R(T_{[m]}^{(X)}T_{[n]}^{(Y)})\nn
&=
g\,\,\chi_{R_1}(T_{[m]}^{(X)})\,\chi_{R_2}(T_{[n]}^{(Y)})\left[
\frac{ \chi_R(T_{2}^{(X,Y)}) }{d_R}-
\frac{ \chi_{R_1}(T_{2}^{(X)}) }{d_{R_1}}-
\frac{ \chi_{R_2}(T_{2}^{(Y)}) }{d_{R_2}}
\right]
\end{align}
But now
\begin{equation}
\chi_{R_1}(T_{[m]})=\left\{
\begin{array}{cl}
(-1)^{c_{R_1}+1}\,(m-1)!\qquad &\text{if $R_1$ is a hook representation}\\
0\qquad &\text{otherwise}
\end{array}
\right.
\end{equation}
where $c_{R_1}$ is the number of boxes in the firs column of the Young diagram associated with the representation $R_1$. A similar equation holds for $\chi_{R_2}(T_{[n]})$. We then have
\begin{equation}\label{chi(R1,R2;R) exp2}
\begin{array}{ll}
&\chi_{R_1,R_2}^R\left( T_{\bar 1,1}T_{[m]}^{(X)}T_{[n]}^{(Y)}\right)=\\[3mm]
&=\left\{
\begin{array}{l}
%g\,\,\chi_{R_1}(T_{[m]}^{(X)})\,\chi_{R_2}(T_{[n]}^{(Y)})\left[
(-1)^{c_{R_1}+c_{R_2}}\,g\,(m-1)!(n-1)!\left[
\frac{ \chi_R(T_{2}^{(X,Y)}) }{d_R}-
\frac{ \chi_{R_1}(T_{2}^{(X)}) }{d_{R_1}}-
\frac{ \chi_{R_2}(T_{2}^{(Y)}) }{d_{R_2}}
\right];\quad R_1,\,R_2 \text{ hooks}\\[3mm]
0\quad \text{otherwise}
\end{array}
\right.
\end{array}
\end{equation}
this is eq. \eqref{chiR full form}.
Let us now restrict to the case in which both $R_1,\,R_2$ are hooks representations. We will denote there representations as $h_1=R_1=(k_1,l_1)$ and $h_2=R_2=(k_2,l_2)$. This also forces the representation $R$ to be at most of depth two, as we derived in Appendix \ref{Appendix: LR rule for hooks}. We now consider such a representation. With the notation given at the beginning of this section, $R=((a_1,b_1),(a_2,b_2))$, it is immediate to write an equation for the normalised character $\frac{\chi_R(T_{2})}{d_R}$
\begin{align}\label{chiR}
\frac{\chi_R(T_{2})}{d_R}&=\frac{1}{2}\sum_i r_i(r_i-2i+1)=a_1(a_1+1)+(a_2+2)(a_2-1)+\\\nonumber
&\qquad\qquad\qquad + 2\sum_{i=3}^{b_2+2}(3-2i)+2\sum_{i=b_2+3}^{b_1+1}(1-i)\nn
&=\frac{1}{2}(a_1^2+a_2^2+a_1+a_2)-\frac{1}{2}(b_1^2+b_2^2+b_1+b_2)\nn
&=\frac{1}{2}(a_1+b_1+1)(a_1-b_1)+\frac{1}{2}(a_2+b_2+1)(a_2-b_2)
\end{align}
We now need the equivalent of this formula for the depth one representations $h_1$ and $h_2$, \textit{i.e.} the hooks. Such an equation can be directly obtained by setting $(a_2,b_2)=(-1,0)$ or $(a_2,b_2)=(0,-1)$ in \eqref{chiR}. We can then write \eqref{chi(R1,R2;R) exp2} as 
\begin{align}\label{chi(R1,R2;R) exp full}
&\chi_{h_1,h_2}^R\left( T_{\bar 1,1}T_{[m]}^{(X)}T_{[n]}^{(Y)}\right)=
\frac{(-1)^{c_{h_1}+c_{h_2}}}{2}\,g\,(m-1)!(n-1)!\times\nn
&\times
\left[\vphantom{\sum}
(a_1+b_1+1)(a_1-b_1)+(a_2+b_2+1)(a_2-b_2)+\right.\\\nonumber
&\qquad\qquad\qquad\left.\vphantom{\sum}
-(k_1+l_1+1)(k_1-l_1)- (k_2+l_2+1)(k_2-l_2)
\right]
\end{align}
where $R=((a_1,b_1),(a_2,b_2))$ and $h_1=(k_1,l_1)$, $h_2=(k_2,l_2)$.

The last piece we need is an equation for the $U(N)$ dimension of a depth two representation $R=((a_1,b_1),(a_2,b_2))$. It is straightforward to write
\begin{align}\label{DimR}
\text{Dim}_N(R)=\frac{(a_1-a_2)(b_1-b_2)}{(a_1+b_2+1)(a_2+b_1+1)}\!
\left(a_1+b_1\atop b_1\right)\!\!\!
\left(a_2+b_2\atop b_2\right)\!\!\!
\left(N+a_1\atop a_1+b_1+1\right)\!\!\!
\left(N+a_2\atop a_2+b_2+1\right)
\end{align}
This equation reduces to its depth 1 equivalent by imposing $(a_2,b_2)=(-1,0)$ or $(a_2,b_2)=(0,-1)$.
It is also helpful to recall the dimension formula for a $S_{l+k+1}$ hook representation $(k,l)$:
\begin{align}\label{dimension eq}
d_R=\left(k+l\atop k\right)
\end{align}

Let us now consider eq. \eqref{char exp}:
\begin{align}\label{char exp sim}
&\la \cO\cO^\dagger\ra=
\frac{1}{m!n!}\,
\sum_{R_1\vdash m\atop R_2\vdash n}\,\sum_{R\vdash m+n}\,
\frac{1}{d_{R_1}\,d_{R_2}\,g}\,\text{Dim}_N(R)\,
\left(\chi_{R_1,R_2}^R\left( T_{\bar 1,1}T_{[m]}^{(X)}T_{[n]}^{(Y)}\right)\right)^2
\end{align}
Inserting eq. \eqref{chi(R1,R2;R) exp full}, \eqref{DimR} and \eqref{dimension eq} into the above equation gives%, recalling that $\cO=\Tr(X^mY^n)$
\begin{align}\label{App: 2pt func}
\langle&\Tr(X^mY^n)\Tr(X^mY^n)^\dagger\rangle\nn
&=\sum_{k_1,l_1=0}^m\,\sum_{k_2,l_2=0}^n\,\,\sum_{a_1,b_1=0\atop a_2,b_2=0}^{n+m}\,g\,\,\delta(k_1+l_1-m)\,\delta(k_2+l_2-n)\,\,\,F(a_1,b_1,a_2,b_2,k_1,l_1,k_2,l_2) 
\end{align}
where we defined the function
\begin{align}\label{F function definition}
&F(a_1,b_1,a_2,b_2,k_1,l_1,k_2,l_2)= 
\frac{k_1! k_2! l_1! l_2!\,(a_1-a_2) (b_1-b_2) }{4(a_1+b_2+1) (a_2+b_1+1)(k_1+l_1+1) (k_2+l_2+1)}\nn
&\times\binom{a_1+b_1}{b_1} \binom{a_2+b_2}{b_2} \binom{N+a_1}{a_1+b_1+1} \binom{N+a_2}{a_2+b_2+1}\times\nn
&\times
((a_1+b_1+1)(a_1-b_1)+(a_2+b_2+1)(a_2-b_2)+\nn
&
\qquad\qquad\qquad
-(k_1+l_1+1)(k_1-l_1)- (k_2+l_2+1)(k_2-l_2))^2
\end{align}

\end{appendix}

\bibliographystyle{utphys}
\bibliography{biblio_Amn}

\end{document}